\documentclass[showpacs,pra,twocolumn]{revtex4}
\usepackage{amsmath}
\usepackage{amsfonts}
\usepackage{amssymb}
\usepackage{amsthm}
\usepackage{graphicx}

\newtheoremstyle{propos}{}{}{}{}{\itshape}{.}{.5em}{}
\theoremstyle{propos}
\newtheorem{Proposition}{Proposition}
\newtheorem{Lemma}{Lemma}

\newcommand{\Tr}{{\textrm{Tr}}}
\newcommand{\Sp}{{\textrm{Sp}}}
\newcommand{\Real}{{\textrm{Re}}}
\begin{document}

\title{Quantifying Superposition}
\author{Johan \AA berg\footnote{Electronic address:
J.Aberg@damtp.cam.ac.uk}}
\affiliation{
Centre for Quantum Computation, Department of
Applied Mathematics and Theoretical Physics, University of Cambridge,
Wilberforce Road, Cambridge CB3 0WA, United Kingdom.}
\date{\today}

\begin{abstract}
Measures are introduced to quantify the degree of superposition in mixed states 
with respect to orthogonal decompositions of the Hilbert space of a quantum system.
 These superposition measures can be regarded as analogues to entanglement
 measures, but can also be put in a more direct relation to the latter. By a second
 quantization of the system it is possible to induce superposition measures from
 entanglement measures. We consider the measures induced from relative entropy
 of entanglement and entanglement of formation. 
We furthermore introduce a class of measures with an operational
 interpretation in terms of interferometry.  We consider the superposition 
 measures under the
 action of subspace preserving and local subspace preserving channels. 
The theory is illustrated with models of an atom undergoing a relaxation process
 in a Mach-Zehnder interferometer.
\end{abstract}

\maketitle

\section{\label{intr} Introduction}
In analogy with entanglement measures we develop means to quantify 
to what degree a mixed quantum state is in superposition with respect 
to given orthogonal subspaces of the Hilbert space of the system. 
We approach this question from two angles. 
In the first approach we pursue the analogies and relations to 
entanglement measures, while in the second we focus 
on the role of superposition in interferometry.  

One specific example of a system in superposition is a single particle in 
superposition between two separated spatial regions.
Such states can display nonlocality 
\cite{Tan, Hardy,Gerry,Bjork,Hessmo} and has been considered as a 
resource for teleportation \cite{telep} and quantum cryptography \cite{crypto}.
In these contexts, the single particle superposition is often not explicitly 
described as such, but is rather modeled using a 
second quantized description of the system, where the superposition 
between the two regions
 is viewed as an entangled state, formed using the vacuum and single 
particle states of the two regions.   
This example suggests that the relation between superposition and 
entanglement goes beyond mere 
analogy, since the former can be 
regarded as the latter, in the above sense.  
With this observation in mind we apply second quantizations in order 
to induce superposition measures from 
entanglement measures, and thus obtain, through the entanglement 
measures, a resource perspective on superposition. 
We investigate the superposition measures induced by relative entropy 
of entanglement \cite{Quaent, entangmes} and entanglement of formation
\cite{mseerr}.

Apart from the fundamental questions concerning superposition, a 
quantitative approach is relevant for the development of techniques 
to probe quantum processes using interferometry  
\cite{Ann, Oi, JA, OiJA, Xiang}. 
In this approach the processes are distinguished by how they affect the 
superposition of the probing particle in the interferometer. A systematic 
investigation of these phenomena thus calls for a quantitative understanding 
of mixed state superposition, which the present investigation may facilitate.
To complement the class of induced superposition measures we consider 
 measures that have a direct operational interpretation in 
terms of interference experiments. 

Lately, a measure has been introduced \cite{Braun} to quantify 
the interference of general quantum processes with respect to a given 
(computational) basis. The superposition measures considered here can 
be regarded as complementary to this approach, 
as we primarily focus on the properties of states,
rather than processes.   

The connection to interferometry suggests a close relation between 
superposition measures and subspace preserving (SP) and local subspace 
preserving (LSP) channels  \cite{Ann}. We analyze the change of superposition 
under the action of SP and LSP channels.
The theory is illustrated with models of an atom that decays to its 
 ground state while propagating through a Mach-Zehnder interferometer. 

The structure of the paper is as follows.
Section \ref{spmes} introduces the concept of superposition measures by two 
specific examples; the relative entropy of superposition and the superposition of 
formation.  In Sec.~\ref{rel}  we relate these measures to the relative entropy of 
entanglement and entanglement of formation. In Sec.~\ref{ind} we introduce the 
concept of induced superposition measures. In Sec.~\ref{indrel} we show 
that relative entropy of superposition is induced by relative entropy of 
entanglement, and in Sec.~\ref{indform} we show that superposition of formation 
is induced by entanglement of formation. 
Section~\ref{unit} introduces another class of superposition measures 
based on unitarily invariant operator norms, and in Sec.~\ref{KyFanMe} 
we consider measures based on Ky-Fan norms.
In Sec.~\ref{inter} it is shown that all Ky-Fan norm based measures have an 
operational interpretation via interferometric measurements, and in 
Sec.~\ref{secbound} we show that the Ky-Fan norm measures are bounded by 
predictability.
Section~\ref{channels} illustrates the relation between superposition measures 
and SP and LSP channels. In Sec.~\ref{chin} we consider the induced 
superposition measures under the action of these classes of channels, and in 
subsection \ref{chno} we similarly consider the superposition measures obtained 
from unitarily invariant norms. In Sec.~\ref{decay} the theory is illustrated with 
models of an atom undergoing relaxation in a Mach-Zehnder interferometer.
The paper is ended with the Conclusions in Sec.~\ref{concl}.

\section{\label{spmes}Superposition measures}
To obtain the simplest possible illustration of the idea of superposition measures, 
consider a two-dimensional Hilbert space spanned by the two orthonormal vectors 
$|1\rangle$ and $|2\rangle$. Assume that we consider 
states $\rho$ for which $\langle 1|\rho|1\rangle = 1/2$, i.e., the probability to find 
the system in either of the two states is $1/2$. All such density operators can be written 
$\rho = (|1\rangle\langle 1| + |2\rangle\langle 2|  +c|1\rangle\langle 2|+
c^{*}|2\rangle\langle 1|)/2$ where the complex number $c$ satisfies $|c|\leq 1$. 
Hence, $c$ is the off-diagonal element when  $\rho$ is represented in the  
$\{|1\rangle, |2\rangle\}$ basis. In one extreme, $|c|=1$,  the state is in a pure equal 
superposition between the two states. In the other extreme, $c=0$, the state is maximally 
mixed and there is no superposition. We can conclude that the off-diagonal element $c$ 
describes the superposition, and it seems intuitively reasonable to take the quantity $|c|$ 
as a measure of ``how much" superposition there is between the two states $|0\rangle$ 
and $|1\rangle$.  This observation generalizes to pairs of orthogonal subspaces, in the 
sense that the off-diagonal block carries the information concerning the superposition.
In Sec.~\ref{unit} we consider superposition measures based directly on this observation, 
while in the following we focus on the analogy with, and relation to, entanglement measures.

Given a finite-dimensional Hilbert space $\mathcal{H}$ we consider a collection of $K$ 
at least one-dimensional  subspaces 
$\boldsymbol{\mathcal{L}} = (\mathcal{L}_{1},\ldots,\mathcal{L}_{K})$ such that 
$\oplus_{k=1}^{K}\mathcal{L}_{k}  =\mathcal{H}$,  i.e., we consider a collection of 
pairwise orthogonal subspaces that spans the entire Hilbert space. When we say that 
$\boldsymbol{\mathcal{L}}$ is a decomposition of $\mathcal{H}$ we assume the above 
mentioned properties.
We let $P_{k}$ denote the projector corresponding to subspace $\mathcal{L}_{k}$, and 
define the channel
\begin{equation}
\Pi(\rho) = \sum_{k=1}^{K}P_{k}\rho P_{k}.
\end{equation}
We write $\Pi_{\boldsymbol{\mathcal{L}}}$ when we want to stress that 
$\Pi$ is defined 
with respect to the decomposition $\boldsymbol{\mathcal{L}}$.
 The effect of $\Pi$ is to remove all the ``off-diagonal blocks" $P_{i}\rho P_{j}$ 
from the density operator, leaving the ``diagonal blocks" $P_{i}\rho P_{i}$ intact.

Superposition measures $A$ are real-valued functions on the set of density 
operators on $\mathcal{H}$, and are defined with respect to some decomposition 
$\boldsymbol{\mathcal{L}}$. To clarify with respect to which decomposition the 
measure is defined we write $A^{\boldsymbol{\mathcal{L}}}$.

The following provides a list of properties that a superposition measure may satisfy.
 Note that we do not claim these properties to be ``necessary conditions" for 
superposition measures, they are merely convenient and reasonable conditions 
that the measures we consider here do satisfy.  For all density operators
 $\rho$ on $\mathcal{H}$,
\begin{itemize}
\item[C1:] $A(\rho)\geq 0$,
\item[C2:] $A(\rho) = 0$ $\Leftrightarrow$ $P_{i}\rho P_{j} = 0$, $\forall ij:i\neq j$,
\item[C3:] $A(U\rho U) = A(\rho)$,  $U = \oplus_{j}U_{j}\nonumber$, where $U_{j}$ 
is unitary on $\mathcal{L}_{j}$,
\item[C4:] $A$ is convex,
\end{itemize}
Property C1 states that the function is non-negative on all states. Condition C2 
says that   $A(\rho)$ is zero if and only if $\rho$ is block-diagonal with respect 
to the decomposition $\boldsymbol{\mathcal{L}}$. Since such a block-diagonal 
state has no superposition between the subspaces in question, C2 thus seems a 
reasonable condition.  The set of block diagonal states can be seen as the analogue 
to the separable states in the context of entanglement (further elaborated in 
Sec.~\ref{ind}). Property C3 states that $A$ is invariant under unitary transformations 
on the subspaces in the decomposition. From an intuitive point of view this seems 
reasonable since such unitary operations do not change the ``magnitude" 
(e.g., some norm) of the off-diagonal operators, and thus should not change the 
``amount" of superposition. This condition is analogous to the invariance of 
entanglement measures under local unitary operations. The last property, C4,  
states that the degree of superposition does not increase under convex 
combinations of states (as is the case for, e.g., $|c|$, as defined above). 

\subsection{Relative entropy of superposition}
We define the \emph{relative entropy of superposition} as
\begin{equation}
\label{forsta}
A_{S}(\rho) = S\boldsymbol{(}\Pi(\rho)\boldsymbol{)}-S(\rho),
\end{equation}
where $S$ denotes the von Neumann entropy $S(\rho) = -\Tr(\rho\ln\rho)$.
Note that
 \begin{equation}
\label{andra}
A_{S}(\rho) = S\boldsymbol{(}\rho||\Pi(\rho)\boldsymbol{)},
\end{equation}
where $S(\rho||\sigma) = \Tr(\rho\ln\rho)- \Tr(\rho\ln\sigma)$, is the relative 
entropy \cite{Umegaki, Lindblad, Wehrl}. One can also show that 
\begin{equation}
\label{tredje}
A_{S}(\rho) = \inf_{\Pi(\sigma)=\sigma}S\boldsymbol{(}\rho||\sigma\boldsymbol{)},
\end{equation}
where the infimum is taken over all block diagonal density operators $\sigma$ 
on $\mathcal{H}$. Hence, an equivalent definition of $A_{S}$ is to minimize 
the relative entropy with respect to all block diagonal states. This may give 
some intuitive understanding of why $A_{S}$ could be regarded as a 
superposition measure.
Note also that $\Pi$ is a mixing enhancing channel \cite{Wehrl} that 
removes off diagonal blocks of the input density operator $\rho$. Intuitively, 
the more ``off diagonal" the density operator $\rho$ is, with respect to the 
subspaces in $\boldsymbol{\mathcal{L}}$, the larger is the difference 
between the two entropies $S\boldsymbol{(}\Pi(\rho)\boldsymbol{)}$ 
and $S(\rho)$. 

By using the properties of relative entropy \cite{Wehrl} one can show 
that $A_{S}$ satisfies conditions C1, C2, C3, and C4.
Moreover, from the additivity and monotonicity of relative entropy 
\cite{Wehrl} it follows that $A_{S}$ is additive and monotone in the 
following sense.
\begin{itemize}
\item Additivity: Let $\boldsymbol{\mathcal{L}}^{(1)}$ be a 
decomposition of $\mathcal{H}^{(1)}$ and let 
$\boldsymbol{\mathcal{L}}^{(2)}$ be a decomposition of $\mathcal{H}^{(2)}$.
If $\rho$ is a density operator on $\mathcal{H}^{(1)}$ and 
$\sigma$ a density operator on $\mathcal{H}^{(2)}$, then
\begin{equation}
A_{S}^{\boldsymbol{\mathcal{L}}^{(1)}\otimes\boldsymbol{\mathcal{L}}^{(2)}}(\rho\otimes\sigma) 
= A_{S}^{\boldsymbol{\mathcal{L}}^{(1)}}(\rho) + A_{S}^{\boldsymbol{\mathcal{L}}^{(2)}}(\sigma).
\end{equation}
\item Monotonicity: Let $\boldsymbol{\mathcal{L}}$ be a decomposition 
of $\mathcal{H}$.
If $\rho$ is a density operator on $\mathcal{H}\otimes\mathcal{H}_{a}$, then
\begin{equation}
A_{S}^{\boldsymbol{\mathcal{L}}}(\Tr_{a}\rho) \leq 
A_{S}^{\boldsymbol{\mathcal{L}}\otimes\hat{1}_{a}}(\rho).
\end{equation}
\end{itemize}

\subsection{Superposition of formation}
In the following, when we say that $(\lambda_{l},|\psi_{l}\rangle)_{l}$ is a 
decomposition of a density operator $\rho$ we intend that all 
$|\psi_{l}\rangle$ are normalized, $\lambda_{l}\geq 0$, and that
\begin{equation}
\label{acdef2}
\rho = \sum_{l}\lambda_{l}|\psi_{l}\rangle\langle\psi_{l}|.
\end{equation}

We define the \emph{superposition of formation} as
\begin{equation}
\label{acdef1}
A_{f}(\rho) = 
\inf\sum_{l}\lambda_{l}S\boldsymbol{(}\Pi(|\psi_{l}\rangle\langle\psi_{l}|)\boldsymbol{)},
\end{equation}
where the infimum is taken with respect to all decompositions 
$(\lambda_{l},|\psi_{l}\rangle)_{l}$ of the density operator $\rho$.
In analogy with entanglement of formation one may note that $A_{f}$ 
is obtained if we minimize  the expected superposition 
(measured with $A_{S}$) needed to prepare $\rho$ from pure state ensembles.

Note that the definition of $A_{f}$ involves an infimum. Technically speaking 
there is a question whether there exists a decomposition 
$(\lambda_{l},|\psi_{l}\rangle)_{l}$  such that 
$A_{f}(\rho) = \sum_{l}\lambda_{l}
S\boldsymbol{(}\Pi(|\psi_{l}\rangle\langle\psi_{l}|)\boldsymbol{)}$. 
To avoid such technicalities we note that for every $\epsilon>0$ there exists a 
decomposition $(\lambda_{l},|\psi_{l}\rangle)_{l}$ of $\rho$  such that 
\begin{equation}
\label{edecompdef}
A_{f}(\rho)\leq  \sum_{l}\lambda_{l}S\boldsymbol{(}\Pi(|\psi_{l}\rangle\langle\psi_{l}|)\boldsymbol{)} 
\leq A_{f}(\rho) +\epsilon.
\end{equation}
We refer to such a decomposition as an \emph{$\epsilon$-decomposition} of $\rho$.

The relative entropy of superposition is bounded from above by the superposition of formation,
\begin{equation}
\label{dzfb}
A_{S}(\rho)\leq A_{f}(\rho).
\end{equation}
To show this we let $(\lambda_{l},|\psi_{l}\rangle)_{l}$ be a $\epsilon$-decomposition 
of $\rho$. The joint convexity of relative entropy \cite{Wehrl}, together with the fact 
that $\Tr\boldsymbol{(}\sigma\ln\Pi(\rho)\boldsymbol{)} = 
\Tr\boldsymbol{(}\Pi(\sigma)\ln\Pi(\rho)\boldsymbol{)}$, can be used to show that
\begin{equation}
A_{S}(\rho)  \leq  
\sum_{l}\lambda_{l}S\boldsymbol{(}\Pi(|\psi_{l}\rangle\langle\psi_{l}|)\boldsymbol{)} 
\leq A_{f}(\rho) +\epsilon.
\end{equation}
If we let $\epsilon\rightarrow 0$, Eq.~(\ref{dzfb}) follows.

Now we shall prove that the superposition of formation  $A_{f}$ satisfies 
conditions C1, C2, C3, and C4.
Condition C1 follows directly from the construction. 
To prove condition C2, assume $A_{f}(\rho)= 0$. From Eq.~(\ref{dzfb}) it 
follows that $A_{S}(\rho)=0$, and we already know that this implies that 
$P_{i}\rho P_{j} =0$, $i,j:i\neq j$. Conversely, if 
$P_{i}\rho P_{j} =0$, $i,j:i\neq j$ it follows that we can find a decomposition 
of $\rho$ where each vector is localized in one of the subspaces 
$\mathcal{L}_{k}$, which implies $A_{f}(\rho)=0$.
Concerning property C3, we note that if $U=\oplus_{j}U_{j}$ then 
$\Pi(U|\psi\rangle\langle\psi| U^{\dagger}) = 
U\Pi(|\psi\rangle\langle \psi|) U^{\dagger}$. Since von Neumann entropy 
is invariant under unitary transformations, condition C3 follows.
To prove condition C4, let $\rho = \sum_{n=1}^{N}\mu_{n}\rho_{n}$ 
be a convex combination of density operators. For each $n$ let 
$(\lambda_{l}^{(n)}, |\psi_{l}^{(n)}\rangle)_{l}$  be an 
$\epsilon$-decomposition of $\rho_{n}$. If we use Eq.~(\ref{edecompdef}) 
it follows that 
\begin{eqnarray}
A_{f}(\rho) & \leq  & \sum_{n}\sum_{l}\mu_{n}\lambda_{l}^{(n)}
S\boldsymbol{(}\Pi(|\psi_{l}^{(n)}\rangle\langle\psi_{l}^{(n)}|)\boldsymbol{)}\nonumber\\
& \leq &\sum_{n}\mu_{n}A_{f}(\rho_{n}) +\epsilon.
\end{eqnarray} 
If we now let $\epsilon\rightarrow 0$ we find that $A_{f}$ is convex. 
Thus, $A_{f}$ satisfies condition $C4$.

\begin{itemize}
\item Subadditivity: Let $\boldsymbol{\mathcal{L}}^{(1)}$ be a decomposition 
of $\mathcal{H}^{(1)}$ and let $\boldsymbol{\mathcal{L}}^{(2)}$ be a 
decomposition of $\mathcal{H}^{(2)}$.
If $\rho$ is a density operator on $\mathcal{H}^{(1)}$ and $\sigma$ a 
density operator on $\mathcal{H}^{(2)}$, then
\begin{equation}
\label{subaddAf}
A_{f}^{\boldsymbol{\mathcal{L}}^{(1)}\otimes\boldsymbol{\mathcal{L}}^{(2)}}(\rho\otimes \sigma) 
\leq A_{f}^{\boldsymbol{\mathcal{L}}^{(1)}}(\rho) + 
A_{f}^{\boldsymbol{\mathcal{L}}^{(2)}}(\sigma).
\end{equation} 
\item Monotonicity: Let $\rho$ be a density operator on 
$\mathcal{H}\otimes\mathcal{H}_{a}$
\begin{equation}
\label{Afmonoton}
A_{f}^{\boldsymbol{\mathcal{L}}}(\Tr_{a}\rho) \leq 
A_{f}^{\boldsymbol{\mathcal{L}}\otimes\hat{1}_{a}}(\rho).
\end{equation}
\end{itemize}
\textit{Proof}.
To prove the subadditivity in Eq.~(\ref{subaddAf}) we let 
$(\lambda_{l}^{(1)},|\psi_{l}^{(1)}\rangle)_{l}$ and  
$(\lambda_{l'}^{(2)},|\psi_{l'}^{(2)}\rangle)_{l'}$ be $\epsilon$-decompositions 
of $\rho$ and $\sigma$, respectively. It follows that   
$(\lambda_{l}^{(1)}\lambda_{l'}^{(2)},|\psi_{l}^{(1)}\rangle|\psi_{l'}^{(2)}\rangle)_{l,l'}$ 
is a decomposition of $\rho\otimes\sigma$ and hence
\begin{eqnarray}
\label{stesub}
A_{f}^{\boldsymbol{\mathcal{L}}^{(1)}\otimes\boldsymbol{\mathcal{L}}^{(2)}}(\rho\otimes \sigma) 
& \leq & \sum_{l,l'}\lambda_{l}^{(1)}\lambda_{l'}^{(2)}
S\boldsymbol{(}\Pi^{(1)}(|\psi_{l'}^{(1)}\rangle\langle\psi_{l'}^{(1)}|)\nonumber\\
& & \otimes \Pi^{(2)}(|\psi_{l}^{(2)}\rangle\langle\psi_{l}^{(2)}|)\boldsymbol{)}\nonumber\\
&\leq & A_{f}^{\boldsymbol{\mathcal{L}}^{(1)}}(\rho) + 
A_{f}^{\boldsymbol{\mathcal{L}}^{(2)}}(\sigma) + 2\epsilon,
\end{eqnarray}
where we at the second inequality have used the additivity of the von Neumann entropy, 
followed by Eq.~(\ref{edecompdef}).
If we let $\epsilon\rightarrow 0$  in Eq.~(\ref{stesub}) we obtain subadditivity.

Next, we turn to the monotonicity in Eq.~(\ref{Afmonoton}). 
Let $(\lambda_{l},|\psi_{l}\rangle)_{l}$  be an $\epsilon$-decomposition 
of the density operator $\rho$ on $\mathcal{H}\otimes\mathcal{H}_{a}$, then
\begin{eqnarray}
\label{dznb}
A_{f}^{\boldsymbol{\mathcal{L}}\otimes\hat{1}_{a}}(\rho) + \epsilon & \geq & 
\sum_{l}\lambda_{l}S\boldsymbol{(}[\Pi\otimes I_{a}]
(|\psi_{l}\rangle\langle\psi_{l}|)\boldsymbol{)}\nonumber\\
 & = &  \sum_{l}\lambda_{l}H(\boldsymbol{p}^{(l)}),
\end{eqnarray}
where $H$ denotes the Shannon entropy, and where 
\begin{equation}
\boldsymbol{p}^{(l)} = (\langle\psi_{l}|P_{1}\otimes\hat{1}_{a}|\psi_{l}\rangle,
\ldots, \langle\psi_{l}|P_{K}\otimes\hat{1}_{a}|\psi_{l}\rangle).
\end{equation}
Now, consider a Schmidt decomposition
$|\psi_{l}\rangle = \sum_{m}\sqrt{r_{m}^{(l)}}|\chi_{m}^{(l)}\rangle|a_{m}^{(l)}\rangle$, 
where $\{|\chi_{m}^{(l)}\rangle\}_{m}$ is  an orthonormal set in $\mathcal{H}$, 
and $\{|a_{m}^{(l)}\rangle\}_{m}$ is orthonormal in $\mathcal{H}_{a}$. We find that 
\begin{equation}
\Tr(P_{k}\otimes\hat{1}_{a}|\psi_{l}\rangle\langle\psi_{l}|)  = 
\sum_{m} r_{m}^{(l)}\Tr(P_{k}|\chi_{m}^{(l)}\rangle\langle \chi_{m}^{(l)}|).
\end{equation}
Thus, if we let $\boldsymbol{p}^{(l)}_{m} = 
[\Tr(P_{k}|\chi_{m}^{(l)}\rangle\langle \chi_{m}^{(l)}|)]_{k}$, 
it follows that $\boldsymbol{p}^{(l)} = \sum_{m}r_{m}^{(l)}\boldsymbol{p}^{(l)}_{m}$. 
Due to the concavity of the Shannon entropy it follows that 
$H(\boldsymbol{p}^{(l)}) \geq \sum_{m}r_{m}^{(l)}H(\boldsymbol{p}^{(l)}_{m})$. 
If we combine this with Eq.~(\ref{dznb}) we find
\begin{equation}
\label{zbzb}
A_{f}^{\boldsymbol{\mathcal{L}}\otimes\hat{1}_{a}}(\rho) + 
\epsilon \geq  \sum_{l}\lambda_{l}\sum_{m}r_{m}^{(l)}H(\boldsymbol{p}^{(l)}_{m}).
\end{equation}
Since $(\lambda_{l},|\psi_{l}\rangle)_{l}$ is a decomposition of $\rho$ it follows 
that $(\lambda_{l}r_{m}^{(l)}, |\chi_{m}^{(l)}\rangle)_{l,m}$ is a decomposition 
of $\Tr_{a}\rho$. Moreover, 
$S\boldsymbol{(}\Pi(|\chi_{m}^{(l)}\rangle\langle \chi_{m}^{(l)}|)\boldsymbol{)} 
= H(\boldsymbol{p}^{(l)}_{m})$. With this in Eq.~(\ref{zbzb}) we obtain
\begin{eqnarray}
A_{f}^{\boldsymbol{\mathcal{L}}\otimes\hat{1}_{a}}(\rho) + \epsilon &\geq 
& \sum_{l}\sum_{m}\lambda_{l}r_{m}^{(l)}
S\boldsymbol{(}\Pi(|\chi_{m}^{(l)}\rangle\langle \chi_{m}^{(l)}|)\boldsymbol{)}\nonumber\\
& \geq & A_{f}^{\boldsymbol{\mathcal{L}}}(\Tr_{a}\rho). 
\end{eqnarray}
If we let $\epsilon\rightarrow 0$ we obtain the monotonicity.
$\Box$

\section{\label{rel}Relations between superposition and entanglement measures}
The results in the previous sections suggest an analogy between the 
relative entropy of superposition and relative entropy of entanglement,  
and similarly  between superposition of formation and entanglement of formation. 
Here we show that there do exist relations between these measures on 
certain classes of states.
The primary reason why we consider these types of states is that they
arise when we  in Sec.~\ref{ind} consider superposition measures induced by 
entanglement measures.

Consider a decomposition $\{\mathcal{L}^{(1)}_{k}\}_{k=1}^{K}$ of a 
finite-dimensional Hilbert space  $\mathcal{H}^{(1)}$, with corresponding 
projectors $P^{(1)}_{k}$. Similarly, consider a decomposition 
$\{\mathcal{L}^{(2)}_{k}\}_{k=1}^{K}$ of a finite-dimensional Hilbert space 
$\mathcal{H}^{(2)}$, and projectors $P^{(2)}_{k}$. We require 
the same number of subspaces in both collections.

On the total Hilbert space $\mathcal{H}^{(1)}\otimes\mathcal{H}^{(2)}$ 
we consider the following subspace
\begin{equation}
\overline{\mathcal{L}} = \bigoplus_{k=1}^{K}\overline{\mathcal{L}}_{k},
\quad \overline{\mathcal{L}}_{k} = \mathcal{L}^{(1)}_{k}\otimes \mathcal{L}^{(2)}_{k}, 
\end{equation}
i.e., the subspace $\overline{\mathcal{L}}$ is an orthogonal sum of product subspaces.
Moreover we define the following decomposition of $\overline{\mathcal{L}}$,
\begin{equation}
\overline{\boldsymbol{\mathcal{L}}} = \{ \overline{\mathcal{L}}_{k}\}_{k=1}^{K}, 
\end{equation}
we  denote the corresponding projectors as
\begin{equation}
\label{Pdef}
\overline{P} = \sum_{k=1}^{K}\overline{P}_{k},\quad 
\overline{P}_{k} = P^{(1)}_{k}\otimes P^{(2)}_{k}.
\end{equation}

In this section we consider states $\rho$ on 
$\mathcal{H}^{(1)}\otimes\mathcal{H}^{(2)}$ such that
\begin{equation}
\overline{P}\rho\overline{P} = \rho.
\end{equation} 
One may note that every bipartite pure state is of this type, 
due to the Schmidt decomposition. In this case the relevant subspaces 
are one-dimensional and correspond to the elements in the Schmidt decomposition.

\subsection{Relative entropy of entanglement}
The relative entropy of entanglement \cite{entangmes} for a bipartite 
state $\sigma$ on a Hilbert space $\mathcal{H}^{(1)}\otimes\mathcal{H}^{(2)}$ 
is defined as $E_{S}(\sigma) = \inf_{\rho} S(\sigma||\rho)$, where the infimum is 
taken over all separable states 
$\rho = \sum_{k}\lambda_{k}\rho_{k}^{(1)}\otimes\rho_{k}^{(2)}$ with respect 
to two subsystems.
If $\rho$ is a separable state such that $E_{S}(\sigma) = S(\sigma||\rho)$, 
then we say that $\rho$ is a \emph{minimizing} separable state with 
respect to $\sigma$.
\begin{Lemma}
\label{leupp}
Let $\rho$ be a density operator on a Hilbert space $\mathcal{H}$  
and let $(\mathcal{L}_{1}, \mathcal{L}_{2})$ be a decomposition of $\mathcal{H}$. 
Then $\rho$ can be written
\begin{eqnarray}
\label{sigmauppd}
\rho &=& p_{1}\sigma_{1} + p_{2}\sigma_{2}+ 
\sqrt{p_{1}p_{2}}\sqrt{\sigma_{1}}D\sqrt{\sigma_{2}} \nonumber\\
 &&+  \sqrt{p_{1}p_{2}}\sqrt{\sigma_{2}}D^{\dagger}\sqrt{\sigma_{1}},
\end{eqnarray}
where $\sigma_{1}$ and $\sigma_{2}$ are density operators such that 
$P_{1}\sigma_{1} P_{1} = \sigma_{1}$, 
$P_{2}\sigma_{2} P_{2} = \sigma_{2}$, $p_{1},p_{2}\geq 0$, 
$p_{0}+ p_{1}=1$, and $D$ satisfies $DD^{\dagger} \leq\hat{1}$.
\end{Lemma}
Note that $DD^{\dagger}\leq \hat{1}$ if and only if  the largest singular 
value of $D$ is less than or equal to $1$.
To prove Lemma \ref{leupp} one can use Lemma 13 in Ref.~\cite{Ann}, 
or a proof almost identical to the proof of Proposition 1 in Ref.~\cite{Uinf}.
Using Lemma \ref{leupp} it is straightforward to prove the following:

\begin{Lemma}
\label{decompo}
Let $\{\mathcal{L}_{k}\}_{k=1}^{K}$ be a decomposition of a subspace 
$\mathcal{L}$ of $\mathcal{H}$. Let $P$ be the projector onto $\mathcal{L}$.
If $\rho$ is a density operator on $\mathcal{H}$, such that $P\rho P =\rho$, 
then $\sigma$ can be written
\begin{eqnarray}
\rho & = & \sum_{k,k':k\neq k'} 
\sqrt{p_{k}p_{k'}}\sqrt{\sigma_{k}}D^{(kk')}\sqrt{\sigma_{k'}}\nonumber\\
             &  & +\sum_{k=1}^{k}p_{k}\sigma_{k},
\end{eqnarray}
where $\sigma_{k}$ are density operators such that 
$P_{k}\sigma_{k}P_{k} = \sigma_{k}$,  $p_{k}\geq 0$ , $\sum_{k}p_{k}=1$, and 
$D^{(kk')}{D^{(kk')}}^{\dagger}\leq \hat{1}$.
\end{Lemma}
Note that this lemma only gives necessary conditions for $\rho$ to be 
a density operator. It does not give sufficient conditions.

We define the function
\begin{equation}
f_{\rho^{*}}^{\sigma}(x,\rho) = S(\sigma||(1-x)\rho^{*}+x\rho),
\end{equation} 
where $\sigma$, $\rho$, and $\rho^{*}$ are density operators and 
$0\leq x\leq 1$.
In Ref.~\cite{entangmes} it is shown that 
\begin{equation}
\label{njkdf}
\frac{\partial f_{\rho^{*}}^{\sigma}}{\partial x}(0,\rho) = 
1-\int_{0}^{\infty}\Tr[Y_{t}(\rho^{*},\sigma) \rho]dt, 
\end{equation}
where
\begin{equation}
Y_{t}(\rho^{*},\sigma)   =  
(\rho^{*} + t\hat{1})^{-1}\sigma(\rho^{*} + t\hat{1})^{-1}.
\end{equation}
Moreover, the following result, which we rephrase as a lemma, 
is proved in Ref.~\cite{entangmes}. 
\begin{Lemma}
\label{sepcond}
Let $\sigma$ and $\rho^{*}$ be  density operators on $\mathcal{H}$, 
and let $\rho^{*}$ be separable.
If 
\begin{equation}
\label{fgnbf}
\frac{\partial f_{\rho^{*}}^{\sigma}}{\partial x}(0,\rho)\geq 0,
\end{equation}
for all pure product states $\rho$, then $\rho^{*}$ has to be a minimizing 
separable state with respect to $\sigma$.
\end{Lemma}
The fact that it is sufficient to satisfy Eq.~(\ref{fgnbf}) for pure product 
states only, follows from the linearity of the right hand side of 
Eq.~(\ref{njkdf}) with respect to $\rho$. 
Note also that 
$\left|\frac{\partial}{\partial x}f_{\rho^{*}}^{\sigma}(0,\rho)-1\right|\leq 1$ 
implies Eq.~(\ref{fgnbf}). 

\begin{Lemma}
\label{omrhosep}
If $\rho^{*}$ is a minimizing separable state with respect to $\sigma$, then 
\begin{equation}
\label{suff}
\left|\frac{\partial f_{\rho^{*}}^{\sigma}}{\partial x}(0,\rho)-1\right|\leq 1,
\end{equation}
for all separable states $\rho$.
\end{Lemma}
The absolute value in Eq.~(\ref{suff}) implies that two inequalities have 
to be satisfied.
One of the inequalities follows since $\rho^{*}$ minimizes the convex 
function $S(\sigma|\rho)$ among all separable states. The other 
inequality follows from the fact that 
$\Tr[Y_{t}(\rho^{*},\sigma) \rho]\geq 0$.

We also need the following lemma:
\begin{Lemma}
\label{singleprod}
Let $P_{1}$ and $P_{2}$ be projectors onto subspaces of 
$\mathcal{H}_{1}$ and $\mathcal{H}_{2}$, respectively.
Suppose $\sigma$ is a density operator on 
$\mathcal{H}_{1}\otimes\mathcal{H}_{2}$ such that 
$P_{1}\otimes P_{2}\sigma P_{1}\otimes P_{2} = \sigma$.
Then every minimizing separable state $\rho^{*}$ with respect to 
$\sigma$ has to satisfy 
$P_{1}\otimes P_{2}\rho^{*} P_{1}\otimes P_{2} = \rho^{*}$.
\end{Lemma}

\textit{Proof}.
Let $P_{1}^{\perp}=\hat{1}_{1}-P_{1}$ and 
$P_{2}^{\perp}= \hat{1}_{2}-P_{2}$.
Now suppose $\rho^{*}$ is a minimizing separable state with respect 
to $\sigma$, but such that  
$P_{1}\otimes P_{2}\rho^{*} P_{1}\otimes P_{2} \neq  \rho^{*}$.
It follows that  $\Tr(P_{1}\otimes P_{2}\rho^{*}) <1$.
Note also that without loss of generality we may assume 
$\Tr(P_{1}\otimes P_{2}\rho^{*}) > 0$.
Consider the channel
\begin{eqnarray}
\Phi(\rho) & = &  P_{1}\otimes P_{2} \rho P_{1}\otimes P_{2}  + 
P_{1}^{\perp}\otimes P_{2}^{\perp} \rho P_{1}^{\perp}\otimes P_{2}^{\perp}\nonumber\\
& & + P_{1}\otimes P_{2}^{\perp} \rho P_{1}\otimes P_{2}^{\perp}\nonumber\\
& &  + P_{1}^{\perp}\otimes P_{2} \rho P_{1}^{\perp}\otimes P_{2}
\end{eqnarray}
and note that $\Phi(\sigma) =\sigma$.
By using contractivity of relative entropy  
$S\boldsymbol{(}\Phi(\sigma)||\Phi(\rho^{*})\boldsymbol{)} \leq S(\sigma||\rho^{*})$ 
\cite{Lindblad2} together with  $\Phi(\sigma) = \sigma$, one can show that 
$S(\sigma||\rho^{*}) >S(\sigma||\widetilde{\rho}) $, where
\begin{equation}
\widetilde{\rho} = 
\frac{P_{1}\otimes P_{2}\rho^{*} P_{1}\otimes P_{2}}{\Tr(P_{1}\otimes P_{2}\rho^{*})}.
\end{equation}
Note that if since $\rho^{*}$ is separable it follows that $\widetilde{\rho}$ 
is also separable.
Hence we have found that $\rho^{*}$ is not a minimizing separable state. 
This is a contradiction. Hence,  
$P_{1}\otimes P_{2}\rho^{*} P_{1}\otimes P_{2} = \rho^{*}$. This proves the lemma.
$\Box$

\begin{Lemma}
\label{decomporho}
Let $\sigma$ be a density operator on $\mathcal{H}_{1}\otimes\mathcal{H}_{2}$ 
such that $\sigma = \overline{P}\sigma \overline{P}$. Then there exists a 
minimizing separable state $\rho^{*}$ with respect to $\sigma$, such that
\begin{equation}
\label{minim}
\rho^{*} = \sum_{k=1}^{K}\Tr(\overline{P}_{k}\sigma)\rho_{k}^{*}, 
\end{equation}
where $\rho_{k}^{*}$ is a minimizing separable state with respect to 
$\sigma_{k} =  \overline{P}_{k} \sigma \overline{P}_{k}/\Tr(\overline{P}_{k}\sigma)$,
if $\Tr(\overline{P}_{k}\sigma)\neq 0$.  In the case  
$\Tr(\overline{P}_{k}\sigma) =  0$ we let $\rho_{k}^{*}=0$.
\end{Lemma}

\textit{Proof}.
We assume that $\rho^{*}$ is as in Eq.~(\ref{minim}). To show that this is a 
minimizing separable state with respect to $\sigma$ it is sufficient to prove 
that $\rho^{*}$ satisfies Eq.~(\ref{fgnbf}) for all pure product states $\rho$. 
By the assumed form of $\rho^{*}$ it  follows that 
\begin{equation}
\rho^{*} + t\hat{1} = \sum_{k=1}^{K}(p_{k}\rho^{*}_{k} + 
t\overline{P}_{k})+ \sum_{k,k':k \neq k'} t P_{k}^{(1)}\otimes P_{k'}^{(2)},
\end{equation} 
where $p_{k} = \Tr(\overline{P}_{k}\sigma)$.
From this it follows that 
\begin{eqnarray}
\label{invutr}
(\rho^{*} + t\hat{1})^{-1} & = & \sum_{k=1}^{K}(p_{k}\rho^{*}_{k} + 
t\overline{P}_{k})^{\ominus}\nonumber\\
& & +\sum_{k,k':k \neq k'} t^{-1} P_{k}^{(1)}\otimes P_{k'}^{(2)},
\end{eqnarray}
where $X^{\ominus}$ denotes the Moore-Penrose (MP) pseudo inverse 
\cite{LanTis}.
By Lemma \ref{singleprod} it follows that 
$\overline{P}_{k} \rho_{k}^{*}\overline{P}_{k} = \rho_{k}^{*}$. 
Hence,  $\overline{P}_{k}(p_{k}\rho^{*}_{k} + 
t\overline{P}_{k})^{\ominus} \overline{P}_{k} = 
(p_{k}\rho^{*}_{k} + t\overline{P}_{k})^{\ominus}$.

 By using  Eq.~(\ref{invutr}), and 
$\overline{P}(P_{k}^{(1)}\otimes P_{k'}^{(2)}) =
\delta_{kk'}P_{k}^{(1)}\otimes P_{k'}^{(2)}$, together with 
Lemma \ref{decompo}, it follows that  
\begin{eqnarray}
\label{unfn}
Y_{t}(\rho^{*},\sigma) 
 & = & \sum_{k,k'=1}^{K} (p_{k}\rho^{*}_{k} 
+ t\overline{P}_{k})^{\ominus}\sigma (p_{k'}\rho^{*}_{k'} 
+ t\overline{P}_{k'})^{\ominus}\nonumber\\
& = & \sum_{k=1}^{K}(p_{k}\rho^{*}_{k} 
+ t \overline{P}_{k})^{\ominus}p_{k}\sigma_{k}(p_{k}\rho^{*}_{k} 
+ t \overline{P}_{k})^{\ominus}\nonumber\\
& &+\sum_{k,k':k \neq k'}\sqrt{p_{k}p_{k'}}(p_{k}\rho^{*}_{k} 
+ t \overline{P}_{k})^{\ominus}\nonumber\\
& & \times \sqrt{\sigma_{k}}D^{(kk')}\sqrt{\sigma_{k'}}(p_{k'}\rho^{*}_{k'} 
+ t \overline{P}_{k'})^{\ominus}.
\end{eqnarray}
By combining Eqs.~(\ref{njkdf}) and (\ref{unfn}) we obtain
\begin{equation}
\label{klasnld}
\left|1-\frac{\partial f}{\partial x}(0,\rho)\right| \leq 
\sum_{k=1}^{K}|A_{k}| + \sum_{kk':k\neq k'}|A_{kk'}|,
\end{equation}
where
\begin{equation}
\label{sdfgs}
A_{k}  = \int_{0}^{\infty}\Tr[ (p_{k}\rho^{*}_{k} 
+ t \overline{P}_{k})^{\ominus}p_{k}\sigma_{k}(p_{k}\rho^{*}_{k} 
+ t \overline{P}_{k})^{\ominus}\rho]dt,
\end{equation}
\begin{eqnarray}
\label{vspoih}
A_{kk'} & =& \sqrt{p_{k}p_{k'}}\int_{0}^{\infty}
\Tr[(p_{k}\rho^{*}_{k} + t \overline{P}_{k})^{\ominus}\sqrt{\sigma_{k}}\nonumber\\
&&\times D^{(kk')}\sqrt{\sigma_{k'}}(p_{k'}\rho^{*}_{k'} 
+ t \overline{P}_{k'})^{\ominus}\rho]dt.
\end{eqnarray}
By the change of variables $t = p_{k}s$ Eq.~(\ref{sdfgs}) can be rewritten as
\begin{equation}
\label{asadfh}
A_{k}   = \Tr(\overline{P}_{k}\rho)\int_{0}^{\infty}
\Tr[ (\rho^{*}_{k} + s \overline{P}_{k})^{\ominus}\sigma_{k}(\rho^{*}_{k} 
+ s \overline{P}_{k})^{\ominus}\overline{\rho}_{k}]ds,
\end{equation}
where
\begin{equation}
\overline{\rho}_{k} = \left\{\begin{array}{cc} 
\overline{P}_{k}\rho\overline{P}_{k}/\Tr(\overline{P}_{k}\rho), & 
\Tr(\overline{P}_{k}\rho)\neq 0,\\
\hat{0},& \Tr(\overline{P}_{k}\rho) =  0.
\end{array}\right.
\end{equation}
Note that if $\rho$ is separable, then $\overline{\rho}_{k}$ is separable. 
From Eqs.~(\ref{asadfh})  and (\ref{njkdf}) it follows that
\begin{equation}
\label{Asl}
|A_{k}| =   \Tr(\overline{P}_{k}\rho)\left|1- \frac{\partial}{\partial x}
f_{\rho_{k}^{*}}^{\sigma_{k}}(0,\overline{\rho}_{k})\right|\leq \Tr(\overline{P}_{k}\rho),
\end{equation}
where the inequality follows from Lemma \ref{omrhosep} and the 
fact that $\overline{\rho}_{k}$ is separable.

We now turn to the $A_{kk'}$ term in Eq.~(\ref{vspoih}).
We make a singular value decomposition \cite{LanTis} of the operator 
$D^{(kk')}$, i.e., there exist orthonormal bases $\{|\alpha:n^{(kk')} \rangle \}_{n}$ 
and $\{|\beta:n^{(kk')} \rangle \}_{n}$ such that 
\begin{equation}
\label{singdec}
D^{(kk')} = \sum_{n}r_{n}^{(kk')}|\alpha:n^{(kk')} \rangle\langle \beta:n^{(kk')}|,
\end{equation}
where $r_{n}^{(kk')}\geq0$. (For convenience we allow singular values with value zero.) 
Since $D^{(kk')}D^{(kk')\dagger}\leq\hat{1}_{kk'}$ it follows that 
$r_{n}^{(kk')}\leq 1$. We assume $\rho$ to be a pure and separable state 
$\rho = |\psi\rangle\langle\psi|$, with $|\psi\rangle =|\psi_{1}\rangle|\psi_{2}\rangle$.
We use this to find
\begin{equation}
\label{intekv}
|A_{kk'}| \leq \sum_{n}\left|\int_{0}^{\infty}R_{n}^{(kk')}(t){Q^{(kk')}_{k:n}}^{*}(t)dt\right|,
\end{equation}
where
\begin{eqnarray}
\label{QRdef}
Q_{n}^{(kk')}(t)  & = & \sqrt{p_{k}}\langle \alpha:n^{(kk')}|\sqrt{\sigma_{k}}(p_{k}\rho^{*}_{k} 
+ t \overline{P}_{k})^{\ominus}|\psi\rangle.\nonumber\\
R_{n}^{(kk')}(t)  & = & \sqrt{p_{k'}}\langle \beta:n^{(kk')}|\sqrt{\sigma_{k'}}(p_{k'}\rho^{*}_{k'} 
+ t \overline{P}_{k'})^{\ominus}|\psi\rangle.\nonumber\\
\end{eqnarray}
We note that $Q_{n}^{(kk')}$ and $R_{n}^{(kk')}$ are $L^{2}(0,\infty)$ functions 
(which follows from Eq.~(\ref{vsnvd1})).
Thus the Cauchy-Schwarz inequality is applicable to the right hand side of Eq.~(\ref{intekv}). 
On the  result of this first application of the Cauchy-Schwartz inequality, 
$\sum_{n}[\int_{0}^{\infty}|R_{n}^{(kk')}(t)|^{2}dt]^{1/2}[\int_{0}^{\infty}|Q_{n}^{(kk')}(t')|^{2}dt']^{1/2}$, 
we  again apply the Cauchy-Schwarz inequality, but this time on this expression regarded as an inner 
product of two finite vectors. 
This results in the upper bound
\begin{eqnarray}
\label{A3nsdk}
|A_{kk'}| &\leq &\sqrt{\sum_{n}\int_{0}^{\infty}|R_{n}^{(kk')}(t)|^{2}dt}\nonumber\\
& & \times\sqrt{\sum_{n'}\int_{0}^{\infty}|Q_{n'}^{(kk')}(t')|^{2}dt'}.
\end{eqnarray}
 Note that
$P_{\alpha}^{(kk')} = \sum_{n}|\alpha:n^{(kk')}\rangle\langle \alpha:n^{(kk')}|$
is a projection operator.
Now we use Eq.~(\ref{QRdef}) and $P_{\alpha}^{(kk')}\leq \hat{1}$, to find
\begin{eqnarray*}
\sum_{n}\int_{0}^{\infty}|Q_{n}^{(kk')}(t)|^{2}dt   & = &  p_{k}
\int_{0}^{\infty} \Tr[(p_{k}\rho^{*}_{k} + t \overline{P}_{k})^{\ominus}\nonumber\\
& & \times \sigma_{k}(p_{k}\rho^{*}_{k} + t \overline{P}_{k})^{\ominus}\rho]dt.
\end{eqnarray*}
On the right hand side of the above equation we recognize $A_{k}$, and thus 
by Eq.~(\ref{Asl}) it follows that
\begin{equation}
\label{vsnvd1}
\sum_{n}\int_{0}^{\infty}|Q_{n}^{(kk')}(t)|^{2}dt  \leq \Tr(\overline{P}_{k}\rho).
\end{equation}
By an analogous reasoning we find that 
$\sum_{n}\int_{0}^{\infty}|R_{n}^{(kk')}(t)|^{2}dt \leq \Tr(\overline{P}_{k'}\rho)$.
We combine the above equations with Eqs.~(\ref{klasnld}), (\ref{Asl}), 
and (\ref{A3nsdk}) we obtain
\begin{equation}
|1-\frac{\partial f}{\partial x}(0,\rho)| \leq 
\left(\sum_{k=1}^{K}\sqrt{\Tr(\overline{P}_{k}\rho)} \right)^{2}.
\end{equation}
Now we again make use of the assumption that  $\rho$ is a (pure) product state 
$\rho=\rho_{1}\otimes\rho_{2}$, and that  
$\overline{P}_{k}=P_{k}^{(1)}\otimes P_{k}^{(2)}$, to show that 
\begin{eqnarray}
\label{bdkmij}
\left|1-\frac{\partial f}{\partial x}(0,\rho)\right| &=&
\left(\sum_{k=1}^{K}\sqrt{\Tr(P_{k}^{(1)}\rho_{1})}
\sqrt{\Tr(P_{k}^{(2)}\rho_{2})}\right)^{2} \nonumber\\
&\leq& 1,
\end{eqnarray}
where we in the last inequality have used the Cauchy-Schwartz inequality. 
According to Lemma \ref{sepcond} it follows that $\rho^{*}$ is a minimizing 
separable state with respect to $\sigma$.
This proves the lemma.
$\Box$

\begin{Proposition}
\label{decompoES}
\label{mainre}
If $\sigma$ is a density operator such that 
$\overline{P}\sigma \overline{P} = \sigma$, then
\begin{eqnarray}
\label{ESuppdelning}
E_{S}(\sigma) & =&\sum_{k}\Tr(\overline{P}_{k}\sigma)
E_{S}\left(\frac{\overline{P}_{k} \sigma \overline{P}_{k}}
{\Tr(\overline{P}_{k}\sigma)}\right) \nonumber\\
& &+ A_{S}^{\overline{\boldsymbol{\mathcal{L}}}}(\sigma).
\end{eqnarray}
In case $\Tr(\overline{P}_{k}\sigma)=0$, the corresponding term 
in the above sum is zero.
\end{Proposition}

\textit{Proof}.
Since $\overline{P}\sigma\overline{P} = \sigma$ it follows by 
Lemma \ref{decomporho} that there exists a minimizing separable 
state $\rho^{*}$ as in Eq.~\ref{minim}, and hence 
$E_{S}(\sigma)  =  S(\sigma||\rho^{*})$. 
Now we can use the properties of $\rho^{*}$ as defined in 
Lemma \ref{decomporho}, noting that 
$\overline{P}_{k}\rho_{k}^{*}\overline{P}_{k} =\rho_{k}^{*}$, 
to calculate $S(\sigma||\rho^{*})$ to be the right hand side of 
Eq.~(\ref{ESuppdelning}). This proves the proposition.
$\Box$ 

With respect to the here considered class of states Proposition 
\ref{decompoES} simplifies the calculation of the relative entropy 
of entanglement by breaking down the problem to the smaller 
subspaces $\overline{\mathcal{L}}_{k}$. If the marginal states 
$\rho_{k}$ are separable, Proposition \ref{decompoES} thus 
provides a closed expression for the relative entropy of entanglement 
of the total state. The following gives a  noteworthy special case. 
\begin{Proposition}
\label{eentropicspec}
If $\overline{P}\sigma\overline{P} =\sigma$, and if at least one of 
the subspaces in each pair 
$(\mathcal{L}^{(1)}_{k},\mathcal{L}^{(2)}_{k})$ is one-dimensional, then
\begin{equation}
E_{S}(\sigma) =  A_{S}^{\overline{\boldsymbol{\mathcal{L}}}}(\sigma).
\end{equation}
\end{Proposition}
This proposition follows directly from Proposition \ref{decompoES} 
since one of the subspaces $\mathcal{L}_{k}^{(1)}$ and  
$\mathcal{L}_{k}^{(2)}$ is one-dimensional, and hence 
$\overline{P}_{k}\sigma\overline{P}_{k}/\Tr(\overline{P}_{k}\sigma)$ 
necessarily is a product state, and thus has zero entanglement.

\subsection{Entanglement of formation}
The entanglement of formation \cite{mseerr} of a density operator 
$\rho$ on $\mathcal{H} ^{(1)}\otimes\mathcal{H}^{(2)}$ is defined as
\begin{equation}
E_{f}(\rho) = \inf\mathcal{S}[(\lambda_{l}, |\psi_{l}\rangle)_{l}],
\end{equation}
where the infimum is taken over all decompositions 
$(\lambda_{l}, |\psi_{l}\rangle)_{l}$ of $\rho$, and where
\begin{equation}
\mathcal{S}[(\lambda_{l}, |\psi_{l}\rangle)_{l}] \equiv 
\sum_{l}\lambda_{l}S\boldsymbol{(}\Tr_{2}(|\psi_{l}\rangle\langle\psi_{l}|)\boldsymbol{)}.
\end{equation}

\begin{Proposition}
\label{partial}
If $\rho$ is a density operator such that 
$\overline{P}\rho\overline{P} = \rho$ then
\begin{eqnarray}
\label{dykab}
 E_{f}(\rho) &\geq  &\sum_{k}\Tr(\overline{P}_{k}\rho)
E_{f}\left(\frac{\overline{P}_{k}\rho \overline{P}_{k}}{
\Tr(\overline{P}_{k}\rho)}\right)\nonumber\\
& & + A_{f}^{\overline{\boldsymbol{\mathcal{L}}}}(\rho).
\end{eqnarray}
In case $\Tr(\overline{P}_{k}\rho)=0$, the corresponding term in the above sum is zero.
\end{Proposition}
\textit{Proof}.
Consider an arbitrary decomposition 
$(\lambda_{l},|\psi_{l}\rangle)_{l}$ of $\rho$. 
Let $p_{k} = \Tr(\overline{P}_{k}\rho)$, and let 
\begin{equation}
\sigma_{k} =\left\{\begin{array}{cc}
\overline{P}_{k}\rho \overline{P}_{k}/p_{k}, & p_{k} \neq 0,\\
\hat{0}, & p_{k} = 0.
\end{array}\right.
\end{equation}
Similarly, let  $p_{l}^{(k)} = \langle\psi_{l}|\overline{P}_{k}|\psi_{l}\rangle$, and 
\begin{equation}
|\psi_{l}^{(k)}\rangle =
\left\{ \begin{array}{cc} \overline{P}_{k}|\psi_{l}\rangle/\sqrt{p_{l}^{(k)}}, & p_{l}^{(k)} \neq 0,\\
 0,  & p_{l}^{(k)} = 0.
\end{array}\right.
\end{equation}
Now we note that $p_{k} =  \sum_{l}\lambda_{l}p_{l}^{(k)}$. 
Hence, if we define $r_{l}^{(k)} = \lambda_{l} p_{l}^{(k)}/p_{k}$, 
we find that $\sum_{l}r_{l}^{(k)} = 1$. Moreover, we find that 
$(r_{l}^{(k)},|\psi_{l}^{(k)}\rangle)_{l}$ is a decomposition of $\sigma_{k}$. 
Note that since $\overline{P}\rho\overline{P} =\rho$ it follows that 
$\overline{P}|\psi_{l}\rangle\langle\psi_{l}|\overline{P} = |\psi_{l}\rangle\langle\psi_{l}|$ 
for all $l$. The structure of the subspace $\overline{\mathcal{L}}$ implies that 
$\Tr_{2}(|\psi_{l}\rangle\langle\psi_{l}|) = 
\sum_{k}p_{l}^{(k)}\Tr_{2}(|\psi_{l}^{(k)}\rangle\langle\psi_{l}^{(k)}|)$. 
One can furthermore show that 
$S\boldsymbol{(}\Pi(|\psi_{l}\rangle\langle \psi_{l}|)\boldsymbol{)} = 
-\sum_{k}p_{l}^{(k)}\ln p_{l}^{(k)}$. By combining these facts it is possible to show that 
\begin{eqnarray}
\label{sbds}
\mathcal{S}[(\lambda_{l},|\psi_{l}\rangle)_{l}] & = &  
\sum_{k}p_{k} \mathcal{S}[(r_{l}^{(k)},|\psi_{l}^{(k)}\rangle)_{l}] \nonumber \\
& &+ \sum_{l}\lambda_{l}S\boldsymbol{(}\Pi(|\psi_{l}\rangle\langle \psi_{l}|)\boldsymbol{)}. 
\end{eqnarray}
In order to calculate $E_{f}$ we have to find the infimum of 
Eq.~(\ref{sbds}) over all decompositions $(\lambda_{l},|\psi_{l}\rangle)_{l}$ of $\rho$.
Note that all $p_{k}$ are fixed by the choice of  $\rho$.
\begin{eqnarray}
\label{fghk}
E_{f}(\rho) & = &  \inf_{(\lambda_{l},|\psi_{l}\rangle)_{l}}
\Big[ \sum_{k}p_{k}\mathcal{S}[(r_{l}^{(k)},|\psi_{l}^{(k)}\rangle)_{l}] \nonumber\\
& &\quad+ \sum_{l}\lambda_{l}
S\boldsymbol{(}\Pi(|\psi_{l}\rangle\langle \psi_{l}|)\boldsymbol{)}\Big] \nonumber\\
& \geq  &  \sum_{k}p_{k}\inf_{(r^{(k)}_{l},|\psi_{l}^{(k)}\rangle)_{l}}
\mathcal{S}[(r_{l}^{(k)},|\psi_{l}^{(k)}\rangle)_{l}]\nonumber\\
 & & + \inf_{(\lambda_{l},|\psi_{l}\rangle)_{l}}\sum_{l}\lambda_{l}
S\boldsymbol{(}\Pi(|\psi_{l}\rangle\langle \psi_{l}|)\boldsymbol{)} \nonumber\\
& = &\sum_{k}p_{k}E_{f}(\sigma_{k})  + A_{f}(\rho),
\end{eqnarray}
where the infima are taken with respect to the decompositions 
$(\lambda_{l},|\psi_{l}\rangle)_{l}$ of $\rho$, and 
$(r^{(k)}_{l},|\psi_{l}^{(k)}\rangle)_{l}$ of $\sigma_{k}$.
This proves the proposition.
$\Box$ 

Note that it is not clear whether the  inequality in Eq.~(\ref{dykab}) 
can be replaced with an equality,  or if there exist states where the inequality is strict.
One can show, however,  that when $\rho$ is pure, then equality holds in 
Proposition \ref{partial}. 
Another special case when equality holds is given by the following proposition.
\begin{Proposition}
\label{ecreaspec}
If $\rho$ is a density operator such that $\overline{P}\rho\overline{P} =\rho$ 
and if at least one of the subspaces in each pair 
$(\mathcal{L}^{(1)}_{k},\mathcal{L}^{(2)}_{k})$ is one-dimensional,  then
\begin{equation}
E_{f}(\rho) =  A_{f}^{\overline{\boldsymbol{\mathcal{L}}}}(\rho).
\end{equation}
\end{Proposition}

\textit{Proof}.
If we neglect all terms but $A_{f}(\rho)$ on the right hand side of 
Eq.~(\ref{dykab}) we find that $E_{f}(\rho) \geq  A_{f}(\rho)$.  
Let $(\lambda_{l},|\psi_{l}\rangle)_{l}$  be an 
$\epsilon$-decomposition of $\rho$ with respect to $A_{f}$. 
We know that $(\lambda_{l},|\psi_{l}\rangle)_{l}$ satisfies Eq.~(\ref{sbds}),
where $|\psi_{l}^{(k)}\rangle = 
\overline{P}_{k}|\psi_{l}\rangle/||\overline{P}_{k}|\psi_{l}\rangle ||$.
Note that since at least one of the subspaces in each pair 
$(\mathcal{L}_{k}^{(1)},\mathcal{L}_{k}^{(2)})$ is one-dimensional, 
it follows that each $|\psi_{l}^{(k)}\rangle$ is a pure product state. 
Hence, 
$S\boldsymbol{(}\Tr_{2}(|\psi_{l}^{(k)}\rangle\langle\psi_{l}^{(k)}|)\boldsymbol{)} = 0$, 
which we can insert in Eq.~(\ref{sbds}) to find that 
$E_{f}(\rho)\leq \mathcal{S}[(\lambda_{l},|\psi_{l}\rangle)_{l}]\leq 
\sum_{l}\lambda_{l}S\boldsymbol{(}\Pi(|\psi_{l}\rangle\langle \psi_{l}|)\boldsymbol{)} 
\leq A_{f}(\rho) +\epsilon$.
If we now let $\epsilon\rightarrow 0$ we find $E_{f}(\rho) \leq A_{f}(\rho)$.
If we combine this with our earlier finding that $E_{f}(\rho)\geq A_{f}(\rho)$, it follows that
 $E_{f}(\rho) = A_{f}(\rho)$ and the proposition is proved.
$\Box$ 

\section{\label{ind}Induced superposition measures}
In the second quantized description of a quantum system we describe the 
occupation states of an orthonormal basis of the original  ``first quantized" 
Hilbert space $\mathcal{H}$ of the system, i.e., to each element in the 
basis we associate a Hilbert space  that describes the possible pure occupation 
states.
The tensor product of these spaces is the second quantized space 
$\mathcal{F}^{(x)}(\mathcal{H})$,  where the $x$ denotes the type of second 
quantization (bosonic, fermionic). 

In the general case we consider a Hilbert space $\mathcal{H}$ 
and a decomposition $\boldsymbol{\mathcal{L}}$ with $K$ elements.  
The corresponding occupation number representation can be regarded 
as a $K$-fold tensor product of second quantizations of the subspaces, i.e., 
$\mathcal{F} = \otimes_{k=1}^{K}F^{(x)}(\mathcal{L}_{k})$.
Since we only consider single particle states it follows that the type of 
second quantization is irrelevant. Moreover, it follows that it is sufficient 
to restrict the analysis to the following subspace of $\mathcal{F}$. We extend each 
space $\mathcal{L}_{k}$ with a vacuum state 
$\widetilde{\mathcal{L}}_{k} = \mathcal{L}_{k}\oplus\Sp\{|0\rangle\}$ 
and construct the space 
$\widetilde{\mathcal{H}} = \otimes_{k=1}^{K}\widetilde{\mathcal{L}}_{k} 
\subseteq \mathcal{F}(\mathcal{H})$. Since the single particle states are all 
elements of this subspace it suffices if we restrict the analysis to 
$\widetilde{\mathcal{H}}$ .

Let $\{|k:l\rangle\}_{l=1}^{N_{k}}$ be an arbitrary but fixed orthonormal 
basis of subspace $\mathcal{L}_{k}$. 
To describe the transition from $\mathcal{H}$ to $\widetilde{H}$ 
it is convenient to use the following operator 
\begin{equation}
\label{Mdef}
M =  \sum_{k=1}^{K}\sum_{l=1}^{N_{k}}|\widetilde{0}\rangle^{\otimes (k-1)}
|\widetilde{1}_{k:l}\rangle |\widetilde{0}\rangle^{\otimes (K-k)}\langle k:l|,
\end{equation}
where $|\widetilde{0}\rangle$ denotes the vacuum state, and 
$|\widetilde{1}_{k:l}\rangle$ a single particle occupation of mode 
$l$ in subsystem $k$.
If $\rho$ is a density operator on the first quantized space, 
then $M\rho M^{\dagger}$ is a density operator on the second quantized space. 
One can note that $M$ is a linear isometry that maps $\mathcal{H}$ 
to the single-particle subspace $\widetilde{\mathcal{H}}_{\textrm{single}}$ 
of  $\widetilde{\mathcal{H}}$, i.e., 
$M^{\dagger}M = \hat{1}_{\mathcal{H}}$ and 
$MM^{\dagger} = P_{\textrm{single}}$, where  $P_{\textrm{single}}$ 
denotes the projector onto $\widetilde{\mathcal{H}}_{\textrm{single}}$.

As an illustration, consider a single particle that can be in two orthogonal orbitals, 
represented by the orthonormal vectors $|1\rangle$ and $|2\rangle$. 
The mapping in Eq.~(\ref{Mdef}) takes the form 
$M = |\widetilde{1}_{1}\rangle|\widetilde{0}_{2}\rangle\langle 1| + 
|\widetilde{0}_{1}\rangle|\widetilde{1}_{2}\rangle\langle 2|$.  
Hence, $|1\rangle$ and $|2\rangle$ are mapped to the product 
states $|\widetilde{1}_{1}\rangle|\widetilde{0}_{2}\rangle$ and 
$|\widetilde{0}_{1}\rangle|\widetilde{1}_{2}\rangle$.
A pure superposition $\alpha |1\rangle + \beta|2\rangle$ 
is mapped to the pure entangled state 
$\alpha |\widetilde{1}\rangle|\widetilde{0}\rangle + 
\beta|\widetilde{0}\rangle|\widetilde{1}\rangle$.  
Finally, a diagonal state $\lambda_{1}|1\rangle\langle 1|+
\lambda_{2}|2\rangle\langle 2|$ is mapped to the separable state 
$\lambda_{1}|\widetilde{1}\rangle \langle \widetilde{1}|\otimes 
|\widetilde{0}\rangle \langle \widetilde{0}| +
\lambda_{2}|\widetilde{0}\rangle \langle \widetilde{0}|\otimes 
|\widetilde{1}\rangle \langle \widetilde{1}|$.
This correspondence can be shown to hold also in the more general case.
Given a density operator $\rho$ on $\mathcal{H}$ and a $K$-element 
decomposition $\boldsymbol{\mathcal{L}}$ of $\mathcal{H}$, the following holds
\begin{itemize}
\item $\rho$ is localized in one subspace, i.e., $P_{j}\rho P_{j} = \rho$ for some $j$, 
if and only if $M\rho M^{\dagger}$ can be written as a product state 
$\widetilde{\rho}_{1}\otimes \ldots\otimes\widetilde{\rho}_{K}$
with respect to all $K$ subsystems.
\item $\rho$ is block diagonal with respect to $\boldsymbol{\mathcal{L}}$, i.e., 
$\Pi_{\boldsymbol{\mathcal{L}}}(\rho) =\rho$, if and only if $M\rho M^{\dagger}$ 
is completely disentangled with respect to all $K$ subsystems, i.e., 
that it can be written as  a convex combination of $K$-fold tensor products of density operators.
\end{itemize}
We are now ready to define induced superposition measures.
Let $M$ be defined by 
Eq.~(\ref{Mdef}) with respect to a $K$-fold orthogonal decomposition 
of a Hilbert space $\mathcal{H}$, and let $E$ be an $K$-partite entanglement measure. 
Define 
\begin{equation}
A(\rho) = E(M\rho M^{\dagger}),
\end{equation}
for all density operators $\rho$ on $\mathcal{H}$.
We say that $A$ is the superposition measure induced by the entanglement 
measure $E$.

At first sight this definition may seem problematic since the operator $M$ 
is not unique. $M$ depends on arbitrary choices of orthonormal bases 
in the subspaces of the first quantized spaces. Thus we may construct 
$\widetilde{M} = MU$ where $U=\oplus_{k=1}^{K}U_{k}$ that would 
give rise to  a new superposition measure. However, if the entanglement 
measure is invariant under local unitary transformations (often regarded as 
a requirement for a ``good" entanglement measure \cite{Quaent}), it can be shown 
that the induced superposition measure is invariant under the various 
choices of $M$.

Note that if $A$ is induced by an entanglement measure $E$, then $A$ 
satisfies C1 if $E(\rho)\geq$ for all density operators $\rho$. 
Moreover, $A$ satisfies C2 if $E$ is such that $E(M\rho M^{\dagger}) =0$ 
if and only if $M\rho M^{\dagger}$ is a completely disentangled state.
Finally, $A$ satisfies C3 if $E$ invariant under local unitary operations, and $A$ 
satisfies C4 if $E$ is convex on the single particle subspace.
\subsection{\label{indrel}Measure induced by relative entropy of entanglement}
 For two complementary subspaces $\mathcal{L}_{1}\oplus\mathcal{L}_{2} = 
\mathcal{H}$ the mapping $M$ maps 
$\mathcal{H}$ to the (single particle) subspace  
$\overline{\mathcal{L}} = \mathcal{L}_{1}\otimes\Sp\{|\widetilde{0}_{2}\rangle\}
\oplus \Sp\{|\widetilde{0}_{1}\rangle\}\otimes\mathcal{L}_{2}$ of 
$\widetilde{H} = \widetilde{\mathcal{L}}_{1}\otimes \widetilde{\mathcal{L}}_{2}$. 
One can see that $\overline{\mathcal{L}}$ is a special case of the class 
of subspaces considered in Sec.~\ref{rel}. We let $\overline{P}$ 
denote the projector onto the subspace $\overline{\mathcal{L}}$.

\begin{Proposition}
With respect to a decomposition 
$(\mathcal{L}_{1},\mathcal{L}_{2})$ of $\mathcal{H}$
the superposition measure induced by the bipartite relative 
entropy of entanglement $E_{S}$  is the relative entropy of 
superposition  $A_{S}$.
\end{Proposition}

\textit{Proof}.
First, we note that the elements in the decomposition 
$\overline{\boldsymbol{\mathcal{L}}} = 
(\mathcal{L}_{1}\otimes\Sp\{|\widetilde{0}_{2}\rangle\}, 
\Sp\{|\widetilde{0}_{1}\rangle\}\otimes\mathcal{L}_{2})$ are such that at 
least one of the subspaces in each product subspace is one-dimensional. 
Next, we note that 
$\overline{P}M\rho M^{\dagger}\overline{P} = M\rho M^{\dagger}$. 
Thus, Proposition \ref{eentropicspec} is applicable, and we can conclude 
that $E_{S}(M\rho M^{\dagger}) = 
A_{S}^{\overline{\boldsymbol{\mathcal{L}}}}(M\rho M^{\dagger})$. It
remains to show that  
$A_{S}^{\overline{\boldsymbol{\mathcal{L}}}}(M\rho M^{\dagger})
=A_{S}^{\boldsymbol{\mathcal{L}}}(\rho)$. 
To prove this we first note that
\begin{equation}
\label{nvk}
S(M\rho M^{\dagger}) = S(\rho),
\end{equation}
which holds since $M$ is a linear isometry and thus preserves the 
eigenvalues of $\rho$.
Moreover, one can show that
$\Pi_{\overline{\boldsymbol{\mathcal{L}}}}(M\rho M^{\dagger}) 
= M\Pi_{\boldsymbol{\mathcal{L}}}(\rho)M^{\dagger}$.
If we combine this with Eq.~(\ref{nvk}) we find
\begin{equation}
\label{Spitransf}
S\boldsymbol{(}\Pi_{\overline{\boldsymbol{\mathcal{L}}}}
(M\rho M^{\dagger})\boldsymbol{)} = 
S\boldsymbol{(}\Pi_{\boldsymbol{\mathcal{L}}}(\rho)\boldsymbol{)}.
\end{equation}
If we combine the definition of $A_{S}$ with Eqs.~(\ref{nvk}) 
and (\ref{Spitransf}) we find that 
$A_{S}^{\overline{\boldsymbol{\mathcal{L}}}}(M\rho M^{\dagger})
=A_{S}^{\boldsymbol{\mathcal{L}}}(\rho)$.
$\Box$

\subsection{\label{indform}Measure induced by entanglement of formation}

\begin{Proposition}
With respect to a decomposition 
$(\mathcal{L}_{1},\mathcal{L}_{2})$ of $\mathcal{H}$
the superposition measure induced by the bipartite 
entanglement of formation $E_{f}$  is the superposition 
of formation  $A_{f}$.
\end{Proposition}

{\it Proof}.
The proof is analogous with the proof of the previous proposition, 
upon the use of Proposition \ref{ecreaspec}, and up to the point 
where we have to prove that 
$A_{f}^{\overline{\boldsymbol{\mathcal{L}}}}(M\rho M^{\dagger})
=A_{f}^{\boldsymbol{\mathcal{L}}}(\rho)$. 
If $(\lambda_{l},|\psi_{l}\rangle)_{l}$ is a decomposition of $\rho$, 
then $(\lambda_{l}, |\widetilde{\psi}_{l}\rangle)_{l}$, 
with $|\widetilde{\psi}_{l}\rangle = M|\psi_{l}\rangle$,  
is a decomposition of $M\rho M^{\dagger}$. Vice versa, 
given a decomposition $(\lambda_{l}, |\widetilde{\psi}_{l}\rangle)_{l}$ 
of  $M\rho M^{\dagger}$, it follows that 
$(\lambda_{l}, |\widetilde{\psi}_{l}\rangle)_{l}$ with 
$|\psi_{l}\rangle = M^{\dagger}|\widetilde{\psi}_{l}\rangle$  
is a decomposition of $\rho$. Moreover, due to Eq.~(\ref{Spitransf}) 
we find that
\begin{equation}
 \sum_{l}\lambda_{l}S\boldsymbol{(}
\Pi_{\overline{\boldsymbol{\mathcal{L}}}}
(|\widetilde{\psi}_{l}\rangle \langle\widetilde{\psi}_{l}|)\boldsymbol{)}
=  \sum_{l}\lambda_{l}S\boldsymbol{(}
\Pi_{\boldsymbol{\mathcal{L}}}(|\psi_{l}\rangle \langle\psi_{l}|)\boldsymbol{)}.
\end{equation}
It follows that 
$A_{f}^{\overline{\boldsymbol{\mathcal{L}}}}(M\rho M^{\dagger})
=A_{f}^{\boldsymbol{\mathcal{L}}}(\rho)$.
$\Box$

The induced superposition measure inherit many of the properties of 
the corresponding entanglement measure. This may at first sight 
seem as an  alternative technique to prove the properties of 
$A_{S}$ and $A_{f}$  proved in Sec.~\ref{spmes}. However, since 
we here only consider bipartite entanglement measures, it means 
that the induced superposition measures are defined only with 
respect to pairs of orthogonal subspaces. Thus, the derivations 
in Sec.~\ref{spmes} are more general. 

We furthermore note that we can induce superposition measures form 
any other entanglement measure. 
One can for example show that the superposition measure induced from entanglement
 cost \cite{cost1} satisfies the condition $C1$, $C2$, $C3$, and $C4$.

\section{\label{unit}Superposition measures from unitarily invariant norms}
Given a decomposition 
$\boldsymbol{\mathcal{L}} = (\mathcal{L}_{1}, \mathcal{L}_{2})$ 
of a Hilbert space $\mathcal{H}$, a density operator $\rho$  on 
$\mathcal{H}$ can be decomposed into two diagonal operators  
$P_{1}\rho P_{1}$ and $P_{2}\rho P_{2}$, and two off-diagonal operators 
$P_{1}\rho P_{2}$ and $P_{2}\rho P_{1}$.
In some sense the off-diagonal operators describes the superposition 
between the two subspaces.
It thus seems reasonable to quantify the ``amount" of superposition 
between the two subspaces by some measure of the magnitude of 
the off-diagonal operator.
We consider superposition measures $A_{u}(\rho) = ||P_{1}\rho P_{2}||$, 
where $||\cdot||$ is a norm on the space 
$\mathcal{B}(\mathcal{L}_{2},\mathcal{L}_{1})$ of linear operators from 
$\mathcal{L}_{1}$ to $\mathcal{L}_{2}$.
In order to make the superposition measure invariant under unitary 
transformations within the subspaces we assume that the norm 
$||\cdot||$ is \emph{unitarily invariant} \cite{Bhatia}, i.e.,  such that 
$||U_{2}CU_{1}|| = ||C||$,
for all $C\in\mathcal{B}(\mathcal{L}_{2},\mathcal{L}_{1})$, 
and all unitary operators $U_{1}$ on $\mathcal{L}_{1}$, 
and all unitary $U_{2}$ on $\mathcal{L}_{2}$.
Such superposition measures we refer to as unitarily invariant 
norm measures.
It is straightforward to show that  every unitarily invariant norm 
measure $A_{u}$ satisfies conditions C1, C2, C3, and C4. 

\subsection{\label{KyFanMe}Measures from Ky-Fan norms}
Given an arbitrary operator 
$C\in \mathcal{B}(\mathcal{L}_{2},\mathcal{L}_{1})$ 
consider its singular values \cite{LanTis} ordered in a non-increasing 
sequence $s_{1}^{\downarrow}(C)\geq s_{2}^{\downarrow}(C)\geq 
\ldots \geq s_{N-1}^{\downarrow}(C)\geq  s_{N}^{\downarrow}(C)$, 
where $N= \min[\dim(\mathcal{L}_{1}), \dim(\mathcal{L}_{2})]$.
Due to convenience we allow singular values to be zero.
The Ky-Fan $k$-norms \cite{Bhatia} are defined as
\begin{equation}
||C||_{(k)} = \sum_{l=1}^{k}s_{l}^{\downarrow}(C),\quad 1\leq k\leq N.
\end{equation}
One may note that $||\cdot||_{(1)}$ is equal to the standard operator 
norm $||C||_{(1)}=\sup_{||\psi||=1}||C|\psi\rangle||$. Moreover,  
$||C||_{(N)} = \Tr(\sqrt{CC^{\dagger}})$ is the trace norm. 
In the following we write $||C||_{(\textrm{Tr})}$ when we wish to 
emphasize that we consider the special case of the trace norm.

Using the Ky-Fan norms we define the superposition measures
$A_{(k)}(\rho) = ||P_{1}\rho P_{2}||_{(k)}$. We refer to these as 
Ky-Fan norm measures. We furthermore write $A_{(\textrm{Tr})}$ 
when we want to emphasize that we consider the superposition 
measure from the trace norm.
Since the Ky-Fan norms are unitarily invariant \cite{Bhatia} 
it follows that these superposition  measures forms a subclass of 
the unitarily invariant norm measures.
However, the Ky-Fan norms have a special position among the unitarily 
invariant norms. Let 
$Q,R\in\mathcal{B}(\mathcal{L}_{2},\mathcal{L}_{1})$, 
then  $||Q||_{(k)}\leq ||R||_{(k)}$ for all $k$, if and only if 
$||Q||\leq ||R||$ for all unitarily invariant norms $||\cdot||$ \cite{Bhatia}.
 We can thus conclude the following:
\begin{Proposition}
\label{impl}
Let $\rho$ and $\sigma$ be arbitrary but fixed density 
operators on $\mathcal{H}$. Then
$A_{(k)}(\rho)\leq A_{(k)}(\sigma)$, $\forall k$
if and only if $A_{u}(\rho)\leq A_{u}(\sigma)$,
for all unitarily invariant norm measures $A_{u}$.
\end{Proposition}
\subsection{\label{inter}Interferometric realization of Ky-Fan norm measures}
Here we consider  interferometric techniques to implement all 
Ky-Fan norm measures.
Consider a single particle that propagates in superposition between two modes, 
corresponding to the orthonormal elements $|1\rangle$ and $|2\rangle$. 
This particle also have an internal degree of freedom (e.g., polarization, spin) 
corresponding to the Hilbert space $\mathcal{H}_{I}$ of dimension $N$. 
The total Hilbert space $\mathcal{H} = 
\mathcal{H}_{I}\otimes\mathcal{H}_{s}$, 
can be decomposed into the two orthogonal subspaces 
$\mathcal{H}_{I}\otimes\Sp\{|0\rangle\}$ and  
$\mathcal{H}_{I}\otimes\Sp\{|1\rangle\}$.

We begin with a description of an interferometric procedure to 
obtain the special case of the trace norm measure $A_{(\textrm{Tr})}$.
We let $\rho$ be the total density operator on 
$\mathcal{H}_{I}\otimes\mathcal{H}_{s}$, 
and apply the unitary operator 
$|1\rangle\langle 1|\otimes \hat{1}_{I} + |2\rangle\langle 2|\otimes U$, 
where $U$ is a variable unitary operator on $\mathcal{H}_{I}$. 
After this we apply a 50-50 beam-splitter followed by a measurement 
to obtain the probability that the particle is found in path $1$.  
We find that this probability is 
$p = \frac{1}{2} + \Real\Tr(\langle 1|\rho|2\rangle U^{\dagger})$.
Hence, by varying $U$ until the maximal probability 
$p_{max} = 1/2 + ||\langle 1|\rho|2\rangle||_{(\textrm{Tr})}$ 
is obtained, we can identify 
$A_{(\textrm{Tr})}(\rho) = ||\langle 1|\rho|2\rangle||_{(\textrm{Tr})}
 = 1/2-p_{max}$. (Note that $A_{(\textrm{Tr})}(\rho)\leq 1/2$, 
which follows from Eq.~(\ref{begr}).) 
We have thus found an operational method to obtain $A_{(\textrm{Tr})}$.

\begin{figure}
\includegraphics[width = 8cm]{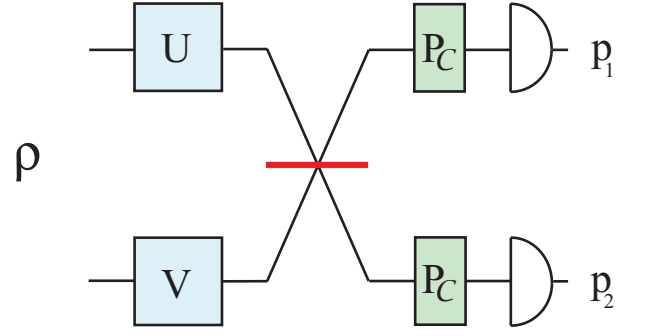}
\caption{\label{fig:interf} (Color online) A procedure to obtain the Ky-Fan 
 norm measure $A_{(k)}= ||\langle 1|\rho|2\rangle||_{(k)}$ by an interferometric technique.
The two paths of the interferometer (upper, lower) corresponds to the 
orthonormal states $|1\rangle$ and $|2\rangle$, spanning the Hilbert 
space $\mathcal{H}_{s}$. The particle has an internal degree of freedom 
(e.g., polarization, spin) represented by the Hilbert space $\mathcal{H}_{I}$. 
The density operator $\rho$ on the total space $\mathcal{H}_{I}\otimes\mathcal{H}_{s}$ 
thus describes both the spatial and internal state of the particle. 
First, two variable unitary operators $U$ and $V$ on $\mathcal{H}_{I}$ are applied, 
one in each path, resulting in the total unitary operator 
$|1\rangle\langle 1|\otimes V + |2\rangle\langle 2|\otimes U$. 
Next, the two paths interfere at a 50-50 beam-splitter.
Let $\mathcal{C}$ be an arbitrary but fixed $k$-dimensional subspace of $\mathcal{H}_{I}$, 
and let $P_{\mathcal{C}}$ be the corresponding projector. 
In both paths we filter with the projector $P_{\mathcal{C}}$, 
i.e., the particle is discarded if its internal state is not found within $\mathcal{C}$. 
Next, we measure the probabilities $p_{1}$ and $p_{2}$ to find the particle in path 
$1$ and $2$, respectively. Note that due to the particle losses in the filtering, 
the two probabilities $p_{1}$ and $p_{2}$ do not in general sum to $1$.
Finally, we vary the unitary operators $U$ and $V$ until we find the maximal 
value of the difference $p_{1}-p_{2}$, and we obtain the desired superposition 
measure as $A_{(k)}(\rho) = \max(p_{1}-p_{2})/2$.}
\end{figure}

In order to obtain the other Ky-Fan norm measures we have to modify 
this scheme  (see figure \ref{fig:interf}).  In this modified scheme 
we first apply two independent unitary operators $U$ and $V$, 
one in each path, 
$|1\rangle\langle 1|\otimes V + |2\rangle\langle 2|\otimes U$, 
after which we apply a 50-50 beam splitter.
Consider next an arbitrary but fixed $k$-dimensional subspace 
$\mathcal{C}$ of $\mathcal{H}_{I}$ and the corresponding projector 
$P_{\mathcal{C}}$. We measure the probability $p_{1}$ to find the 
particle in path $1$ and simultaneously the internal state in subspace 
$\mathcal{C}$, i.e., $p_{1}$ is the expectation value of 
$|1\rangle\langle 1|\otimes P_{\mathcal{C}}$.  
Alternatively, we may filter with the projector $P_{\mathcal{C}}$, 
after which we measure the probability to find the particle in path $1$. 
(Note that we should not normalize after the filtering, i.e., 
we must keep track of the number of particles we have lost.) 
Similarly, we measure the probability $p_{2}$ to find the particle in path $2$ 
and simultaneously in subspace $\mathcal{C}$. 
This results in the probabilities 
$p_{1} = q_{1} + q_{2} + r$ and $p_{2} = q_{1} + q_{2} - r$, where
\begin{eqnarray}
q_{1} &=& \frac{1}{2}\Tr(P_{\mathcal{C}}V\langle 1|\rho|1\rangle V^{\dagger}),
\,\, q_{2} =
\frac{1}{2}\Tr(P_{\mathcal{C}}U\langle 2|\rho| 2\rangle U^{\dagger})\nonumber\\
r & =& \Real\Tr(P_{\mathcal{C}}V \langle 1|\rho|2\rangle  U^{\dagger}),
\end{eqnarray}
and consequently 
$p_{1}-p_{2}=  2r = 2\Real\Tr(P_{\mathcal{C}}V\langle 1|\rho|2\rangle U^{\dagger})$. 
(Note that $p_{1}$ and $p_{2}$ in general do not sum to $1$.) Moreover, 
\begin{equation}
\sup_{U,V}\Real\Tr(P_{\mathcal{C}}V\langle 1|\rho|2\rangle U^{\dagger}) 
= ||\langle 1|\rho|2\rangle||_{(k)},
\end{equation}
which can be proved by making a singular value 
decomposition of $\langle 1|\rho|2\rangle$.
Thus, if we vary the unitary operators $U$ and $V$ in such a 
way that we maximize the difference $p_{1}-p_{2}$, the maximal 
value of this quantity is equal to $2A_{(k)}(\rho)$.
Hence, this interferometric technique allows us to obtain all Ky-Fan norm measures.

\subsection{\label{secbound}Bounds from predictability}
In view of Sec.~\ref{inter} the Ky-Fan norm measures can be interpreted 
as generalized visibilities in an interferometric setup. 
In Refs.~\cite{Englert,Durr} it is showed that the visibility (in the ordinary sense) 
is bounded by the predictability of finding the particle in the two paths.  
Here we show that all the Ky-Fan norm measures satisfy similar bounds.
If we let $\mathcal{L}_{1}$ and $\mathcal{L}_{2}$ correspond to the two paths, 
and let $\rho$ be some arbitrary state of the particle, then 
$p_{1} = \Tr(\mathcal{L}_{1}\rho)$ and $p_{2} =\Tr(\mathcal{L}_{2}\rho)$ 
are the probabilities of finding the particle in path $1$ and $2$, respectively, 
and the predictability is $\mathcal{P} = |p_{1}-p_{2}|$.

We first note the following inequality which is a special case of 
Theorem IV.2.5 in \cite{Bhatia}.
If $A$ and $B$ are operators on the finite-dimensional Hilbert 
space $\mathcal{H}$, then
\begin{equation}
\label{bound}
||AB||_{(k)} \leq \sum_{l=1}^{k}s^{\downarrow}_{l}(A)s^{\downarrow}_{l}(B).
\end{equation}

Let $\rho$ be a density operator on $\mathcal{H}$ and define the ``marginal" 
density operators 
$\sigma_{n} = \mathcal{L}_{n}\rho\mathcal{L}_{n}/p_{n}$, $n= 1,2$. 
(In case $p_{n}=0$, then we let $\sigma_{n}$ be the zero operator.)  
Given these probabilities and marginal states we find that the Ky-Fan 
norm measures satisfy the bound
\begin{equation}
\label{begr}
A_{(k)}(\rho) \leq 
\sqrt{p_{1}p_{2}}\sum_{l=1}^{k}\sqrt{\lambda_{l}^{\downarrow}
(\sigma_{1})}\sqrt{\lambda_{l}^{\downarrow}(\sigma_{2})},
\end{equation}
where $\lambda_{l}^{\downarrow}(\cdot)$ denotes the eigenvalues of 
the enclosed operator ordered nonincreasingly. 
Note that if the dimension of one of the subspaces is strictly larger than 
the other, we may add zeros to the vector with fewer eigenvalues.
To prove the bound in Eq.~(\ref{begr}) we first combine 
Lemma  \ref{leupp} with Eq.~(\ref{bound}), 
to find that $||P_{1}\rho P_{2}||_{(k)}\leq 
\sqrt{p_{1}p_{2}}\sum_{l}s^{\downarrow}_{l}(\sqrt{\sigma_{1}})
s_{l}^{\downarrow}(D\sqrt{\sigma_{2}})$. 
Next, we use that $s_{l}^{\downarrow}(D\sqrt{\sigma_{2}})\leq 
||D||_{(1)}s_{l}^{\downarrow}(\sqrt{\sigma_{2}})$ (see, e.g., \cite{Bhatia}) 
combined with $||D||_{(1)}\leq 1$. Finally, we use the fact that 
$\sigma_{1}$ and $\sigma_{2}$ are positive semi-definite, 
and thus $s_{l}^{\downarrow}(\sigma_{n}^{1/2})  = 
[\lambda_{l}^{\downarrow}(\sigma_{n})]^{1/2}$, 
from which  Eq.~(\ref{begr}) follows. 

Given the probabilities $p_{1},p_{2}$ and the set of eigenvalues 
$\{\lambda_{1,l}^{\downarrow}\}_{l}$ and 
$\{\lambda_{2,j}^{\downarrow}\}_{j}$ of the two marginal operators, 
the bound in Eq.~(\ref{begr}) is sharp, since one obtains equality in 
Eq.~(\ref{begr}) for the density operator 
\begin{eqnarray*}
\rho & = & p_{1}\sum_{l=1}^{L}\lambda_{1,l}^{\downarrow}|1:l\rangle\langle 1:l| 
+ p_{2}\sum_{j=1}^{J}\lambda_{2,j}^{\downarrow}|2:j\rangle\langle 2:j| \nonumber\\
& & + \sqrt{p_{1}p_{2}}\sum_{l=1}^{\min(L,J)}\sqrt{\lambda_{1,l}^{\downarrow}
\lambda_{2,l}^{\downarrow}}( |1:l\rangle\langle 2:l| +|2:l\rangle\langle 1:l|),  
\end{eqnarray*}
where $\{|1:l\rangle\}_{l=1}^{L}$ and $\{|2:j\rangle\}_{j=1}^{J}$ are 
arbitrary orthonormal bases of $\mathcal{L}_{1}$ and 
$\mathcal{L}_{2}$, respectively.

Let us now take a closer look on Eq.~(\ref{begr}). We first note that 
$\sum_{l=1}^{k}\sqrt{\lambda_{l}^{\downarrow}(\sigma_{1})}
\sqrt{\lambda_{l}^{\downarrow}(\sigma_{2})}\leq 1$, and hence 
$A_{(k)}(\rho)\leq \sqrt{p_{1}p_{2}}\leq 1/2$. 
Using this one can show that $A_{(k)}^{2}(\rho) + \mathcal{P}^{2}\leq 1$
 (although $A_{(k)}(\rho)\leq\sqrt{p_{1}p_{2}}$ gives a stronger bound on $A_{(k)}$).
Hence, similarly to the ordinary visibility, the Ky-Fan norm measures are 
bounded by the predictability. We moreover see that the distribution of 
eigenvalues of the two marginal density operators put a limit on the 
superposition measures. For $A_{(\textrm{Tr})}$ the upper bound 
reduces to the classical fidelity between the two distributions of eigenvalues, 
and hence a necessary condition for $A_{(\textrm{Tr})}$ to obtain the 
maximum $\sqrt{p_{1}p_{2}}$ (for $0< p_{1}<1$) is that the eigenvalues 
of the two marginal density operators are equal. For $A_{(k)}$ to attain 
this maximum a necessary condition is that the marginal density operators 
both are of at most rank $k$ and have the same nonzero eigenvalues.  
Hence, for all the Ky-Fan norms to simultaneously attain the maximum 
$\sqrt{p_{1}p_{2}}$, the two marginal states have to be pure. Moreover, 
the total state has to be pure and be possible to write as 
$\sqrt{p_{1}}|\psi_{1}\rangle + \sqrt{p_{1}}|\psi_{2}\rangle$, 
where $|\psi_{1}\rangle$, and $|\psi_{1}\rangle$ are normalized, 
and such that $P_{1}|\psi_{1}\rangle =|\psi_{1}\rangle$ and  
$P_{2}|\psi_{2}\rangle =|\psi_{2}\rangle$.

\section{\label{channels} Subspace preserving and local subspace preserving channels}
With respect to a two-element decomposition 
$(\mathcal{L}_{1},\mathcal{L}_{2})$ of a Hilbert space $\mathcal{H}$ 
a subspace preserving channel is such that 
$\Tr\boldsymbol{(}P_{1}\Phi(\rho)\boldsymbol{)} = \Tr(P_{1}\rho)$ 
for all density operators $\rho$ on $\mathcal{H}$ \cite{Ann}. 
If the decomposition corresponds to the two paths of a Mach-Zehnder interferometer, 
as described in Sec.~\ref{inter}, this condition means that there is no transfer of 
the particle from one path to the other; the probability weights on the two 
paths are preserved under the action of this class of channels.  
Apart from this restriction we can use any type of information transfer or 
sharing of correlated resources between the two paths when we  generate these channels. 
The local subspace preserving channels \cite{Ann} forms the subset of the SP 
channels that can be generated using only local operations at the two paths of the interferometer.
Another way to put this is to say that  
 the LSP channels are those SP channels that can be obtained from product 
channels on the second quantization of $\mathcal{H}$. In terms of the 
mapping $M$ introduced in Sec.~\ref{ind} an SP channel $\Phi$ is 
LSP if there exist a product channel 
$\widetilde{\Phi}_{1}\otimes\widetilde{\Phi}_{2}$ on 
$\widetilde{\mathcal{H}}$ such that 
$\Phi(\rho) = M^{\dagger}[\widetilde{\Phi}_{1}
\otimes\widetilde{\Phi}_{2}](M\rho M^{\dagger})M$. 
A more detailed introduction to these concepts can be found in 
Ref.~\cite{Ann}, and their use in the context of interferometry can 
be found in Refs.~\cite{JA, OiJA}. Here we consider the relation 
between superposition measures and these two classes of channels.

\subsection{\label{chin}Induced measures}
\begin{Proposition}
\label{nessuff}
Let $\boldsymbol{\mathcal{L}}$ be a decomposition of $\mathcal{H}$, 
and suppose $\Phi$ is a channel on $\mathcal{H}$. Then 
\begin{equation}
\label{Asdec}
A_{S}^{\boldsymbol{\mathcal{L}}}\boldsymbol{(}\Phi(\rho)\boldsymbol{)}
\leq A_{S}^{\boldsymbol{\mathcal{L}}}(\rho),
\end{equation}
for all density operators $\rho$ on $\mathcal{H}$, if and only if 
\begin{equation}
\label{compo}
\Pi_{\boldsymbol{\mathcal{L}}}\circ\Phi\circ\Pi_{\boldsymbol{\mathcal{L}}} 
= \Phi\circ\Pi_{\boldsymbol{\mathcal{L}}}.
\end{equation}
\end{Proposition}
Here, ``$\circ$" denotes compositions of mappings. 
In words this proposition states that a channel does not increase the 
relative entropy of superposition (for any input state) if and only if it maps 
all block diagonal states to block diagonal states. One may note a similar 
relation between relative entropy of entanglement and separable channels 
\cite{comment,Rains,BenDiVi,CiDu}.

\textit{Proof}.
For the ``only if" part of the proof we note that $\Pi(\rho)$ is block diagonal, 
and thus (since $A_{S}$ satisfies C2) $A_{S}\boldsymbol{(}\Pi(\rho)\boldsymbol{)}=0$. 
From Eq.~(\ref{Asdec}), it follows that  
$A_{S}\boldsymbol{(}\Phi\circ\Pi(\rho)\boldsymbol{)}=0$.  
Thus, according to condition  C2, it follows that  
$\Phi\circ\Pi(\rho)$ is block-diagonal, i.e., Eq.~(\ref{compo}) follows.

For the ``if" part of the proof we note the trivially satisfied relation 
$S\boldsymbol{(}\Pi\circ\Phi(\rho)||\Pi\circ\Phi\circ\Pi(\rho) \boldsymbol{)}\geq 0$. 
For any density operators $w$ and $\sigma$  it holds that 
$\Tr\boldsymbol{(}w\ln \Pi(\sigma)\boldsymbol{)} 
= \Tr\boldsymbol{(}\Pi(w)\ln\Pi(\sigma)\boldsymbol{)}$. 
Hence, we can rewrite the previous inequality as
$\Tr\boldsymbol{(}\Phi(\rho)\ln \Pi\circ\Phi(\rho)\boldsymbol{)} 
\geq \Tr\boldsymbol{(}\Phi(\rho)\ln\Pi\circ\Phi\circ\Pi(\rho) \boldsymbol{)}$. 
This we combine with Eq.~(\ref{compo}) to obtain 
$\Tr\boldsymbol{(}\Phi(\rho)\ln \Pi\circ\Phi(\rho)\boldsymbol{)}
 \geq \Tr\boldsymbol{(}\Phi(\rho)\ln\Phi\circ\Pi(\rho) \boldsymbol{)}$, 
which can be used to show that $A_{S}(\Phi(\rho)) \leq 
 S\boldsymbol{(}\Phi(\rho)||\Phi\circ\Pi(\rho)\boldsymbol{)}$.
Now we use the contractivity of relative entropy \cite{Lindblad2}, 
which implies that $S\boldsymbol{(}\Phi(\rho)||\Phi\circ\Pi(\rho)\boldsymbol{)}
\leq S\boldsymbol{(}\rho||\Pi(\rho)\boldsymbol{)} = A_{S}(\rho)$. 
Thus follows Eq.~(\ref{Asdec}) and the proposition is proved.
$\Box$

Let $\boldsymbol{\mathcal{L}}= (\mathcal{L}_{1},\mathcal{L}_{2})$ be a 
decomposition of $\mathcal{H}$.
\begin{itemize}
\item For a channel $\Phi$ that is SP with respect to $\boldsymbol{\mathcal{L}}$ 
it holds that $A_{S}^{\boldsymbol{\mathcal{L}}}\boldsymbol{(}\Phi(\rho)\boldsymbol{)}
\leq A_{S}^{\boldsymbol{\mathcal{L}}}(\rho)$, 
for every density operator $\rho$ on $\mathcal{H}$.
\item For every superposition measure $A$ that is induced by a bipartite 
entanglement measure it holds that 
$A^{\boldsymbol{\mathcal{L}}}\boldsymbol{(}\Phi(\rho)\boldsymbol{)}
\leq A^{\boldsymbol{\mathcal{L}}}(\rho)$, for every density operator $\rho$ 
on $\mathcal{H}$, and every channel $\Phi$ that is LSP with respect to 
$\boldsymbol{\mathcal{L}}$.
\end{itemize}
To show that SP channels do not increase the relative entropy of superposition 
we note that if a channel is SP it follows, \cite{Ann} Proposition 4, 
that it can be written
\begin{eqnarray}
\Phi(\rho) = &  P_{1}\Phi(P_{1}\rho P_{1})P_{1} 
+ P_{2}\Phi(P_{2}\rho P_{2})P_{2}\nonumber\\
& + P_{1}\Phi(P_{1}\rho P_{2})P_{2} 
+ P_{2}\Phi(P_{2}\rho P_{1})P_{1}
\end{eqnarray}
This can be used to show that $\Phi$ satisfies Eq.~(\ref{compo}). 
Hence, according to Proposition 
 \ref{nessuff} it follows that $\Phi$ cannot increase the relative entropy of superposition.
To show that LSP channels do not increase any induced superposition measure
we note that since LSP channels correspond to  special types of product 
channels on the second quantized spaces \cite{Ann}. 
It directly follows that the superposition measures induced by entanglement 
measures cannot increase under LSP channels, if we make the natural 
assumption that entanglement 
measures do not increase under local operations.  

\subsection{\label{chno}Norm measures}

\begin{Lemma}
\label{lekontrak}
Let $\{V_{k}\}_{k=1}^{K}$ and $\{W_{l}\}_{l=1}^{L}$ be operators on 
$\mathcal{H}$ be such that $\sum_{k}V_{k}^{\dagger}V_{k}\leq \hat{1}$ 
and $\sum_{l}W_{l}^{\dagger}W_{l}\leq \hat{1}$, and let $\boldsymbol{C}$ 
be a complex $K\times L$ matrix such that $CC^{\dagger}\leq I$. Then,
\begin{equation}
\label{fhjmf}
||\sum_{kl}C_{kl}V_{k}Q W_{l}^{\dagger}||_{(\textrm{Tr})}\leq ||Q||_{(\textrm{Tr})},
\end{equation}
for all linear operators $Q$ on $\mathcal{H}$.
\end{Lemma}
In other words this lemma states that the trace norm of an operator cannot 
increase under this type of transformations.

\textit{Proof}.
We begin with a singular value decomposition 
$\boldsymbol{C} = \boldsymbol{U}^{(a)}\boldsymbol{S}{\boldsymbol{U}^{(b)}}^{\dagger}$, 
where $S_{kl} =s_{k}^{\downarrow}(\boldsymbol{C})\delta_{kl}$ 
for $k,l\leq \min(K,L)$ and $S_{kl} =0$ otherwise. 
We use the unitary matrices $\boldsymbol{U}^{(a)}$ and $\boldsymbol{U}^{(b)}$ 
to define $\overline{V}_{k'} = \sum_{k}V_{k}U_{kk'}^{(a)}$ and  
$\overline{W}_{l'} = \sum_{l}W_{l}U_{ll'}^{(b)}$.
It follows that
\begin{equation}
\label{nrq1}
\sum_{kl}C_{kl}V_{k}Q W_{l}^{\dagger} = 
\sum_{n=1}^{\min(K,L)}s_{n}^{\downarrow}(\boldsymbol{C})\overline{V}_{n}
Q\overline{W}_{n}^{\dagger},
\end{equation}
for all linear operators $Q$ on $\mathcal{H}$. We furthermore find that 
$\sum_{n}\overline{V}_{n}^{\dagger}\overline{V}_{n}\leq \hat{1}$ and 
$\sum_{n}\overline{W}_{n}^{\dagger}\overline{W}_{n}\leq \hat{1}$. One can show that
\begin{equation}
\label{nrq2}
|| \sum_{n}s_{n}^{\downarrow}(\boldsymbol{C})\overline{V}_{n}Q
\overline{W}_{n}^{\dagger} ||_{(\textrm{Tr})} \leq 
\sum_{n} ||\overline{V}_{n}Q\overline{W}_{n}^{\dagger} ||_{(\textrm{Tr})},
\end{equation}
where we have used  that the condition  
$\boldsymbol{C}\boldsymbol{C}^{\dagger}\leq I$ 
implies $s_{n}^{\downarrow}(\boldsymbol{C})\leq 1$. 
Consider now a singular value decomposition 
$Q = \sum_{j}s_{j}^{\downarrow}(Q)|f_{j}\rangle\langle d_{j}|$, 
where $\{|f_{j}\rangle \}_{j}$ and $\{|d_{j}\rangle \}_{j}$ both are orthonormal sets.
\begin{eqnarray}
\label{nrq3}
 ||\overline{V}_{n}Q\overline{W}_{n}^{\dagger} ||_{(\textrm{Tr})} 
& \leq & \sum_{j}s_{j}^{\downarrow}(Q) 
|| \overline{V}_{n}|f_{j}\rangle\langle d_{j}| \overline{W}_{n}^{\dagger}||_{(\textrm{Tr})}\nonumber\\
& = & \sum_{j}s_{j}^{\downarrow}(Q) \sqrt{\langle f_{j}|
 \overline{V}_{n}^{\dagger}\overline{V}_{n}|f_{j}\rangle}\nonumber\\
& & \times\sqrt{\langle d_{j}| \overline{W}_{n}^{\dagger} \overline{W}_{n}| d_{j} \rangle}
\end{eqnarray}
Now we use the Cauchy-Schwartz inequality to find that
$\sum_{n} \langle f_{j}| \overline{V}_{n}^{\dagger}\overline{V}_{n}|f_{j}\rangle^{1/2}
\langle d_{j}| \overline{W}_{n}^{\dagger} \overline{W}_{n}| d_{j} \rangle^{1/2}\leq  1$.
If this is combined with Eqs.~(\ref{nrq1}) - (\ref{nrq3}), and with the fact that 
$||Q||_{(\textrm{Tr})} = \sum_{j}s_{j}^{\downarrow}(Q)$, we find Eq.~(\ref{fhjmf}) 
and have thus proved the lemma.
$\Box$

\begin{Proposition}
\label{nessuffigen}
Let $\boldsymbol{\mathcal{L}} =(\mathcal{L}_{1},\mathcal{L}_{2})$ be a two-element 
decomposition of $\mathcal{H}$, and suppose $\Phi$ is a channel on $\mathcal{H}$. Then 
\begin{equation}
\label{Amdec}
A_{(\textrm{Tr})}^{\boldsymbol{\mathcal{L}}}\boldsymbol{(}\Phi(\rho)\boldsymbol{)}
\leq A_{(\textrm{Tr})}^{\boldsymbol{\mathcal{L}}}(\rho),
\end{equation}
for all density operators $\rho$ on $\mathcal{H}$, if and only if 
\begin{equation}
\label{compoigen}
\Pi_{\boldsymbol{\mathcal{L}}}\circ\Phi\circ\Pi_{\boldsymbol{\mathcal{L}}}= 
\Phi\circ\Pi_{\boldsymbol{\mathcal{L}}}.
\end{equation}
\end{Proposition}
Note that $A_{(\textrm{Tr})}$ denotes the superposition measure obtained 
from the trace norm $||\cdot||_{(\textrm{Tr})}$.

\textit{Proof}.
For the ``only if" part of the proof we note that $\Pi(\rho)$ is block diagonal, 
and thus $A_{(\textrm{Tr})}\boldsymbol{(}\Pi(\rho)\boldsymbol{)}=0$, 
since $A_{(\textrm{Tr})}$ satisfies C2. From Eq.~(\ref{Amdec}), it follows that  
$A_{(\textrm{Tr})}\boldsymbol{(}\Phi\circ\Pi(\rho)\boldsymbol{)}=0$.  
Thus, according to condition C2, it follows that  $\Phi\circ\Pi(\rho)$ is block-diagonal, 
and  Eq.~(\ref{compoigen}) follows.

For the ``if" part of the proof we define the linear map 
$\Pi^{\perp}(Q) = P_{1}QP_{2} + P_{2}QP_{1}$, 
for all linear operators $Q$ on $\mathcal{H}$. 
(Note that $\Pi^{\perp}$ is not a CPM, and not even a positive map.) 
One can show that the condition in Eq.~(\ref{compoigen}) is equivalent to  
$\Pi^{\perp}\circ\Phi\circ\Pi^{\perp} = \Pi^{\perp}\circ\Phi$. 
Now we note that $\Pi^{\perp}$ can be written such that it satisfies the 
conditions in Lemma \ref{lekontrak} (with $\boldsymbol{C}$ the identity matrix and 
$V_{1} = P_{1}$, $V_{2}=P_{2}$, $W_{1} = P_{2}$, and $W_{2} = P_{1}$). 
Moreover, one can note that $\Phi$ (like every other channel) also satisfies the 
conditions in Lemma \ref{lekontrak}. Thus, we let $\rho$ be an arbitrary density 
operator on $\mathcal{H}$, and we use Lemma \ref{lekontrak} twice to obtain 
\begin{eqnarray}
\label{sdfnb}
||\Pi^{\perp}\circ\Phi(\rho)||_{(\textrm{Tr})} & = & 
||\Pi^{\perp}\circ\Phi\circ\Pi^{\perp}(\rho)||_{(\textrm{Tr})} \nonumber\\
& \leq &  ||\Phi\circ\Pi^{\perp}(\rho)||_{(\textrm{Tr})} \nonumber\\
& \leq &  ||\Pi^{\perp}(\rho)||_{(\textrm{Tr})}. 
\end{eqnarray}
Now we note that for each Hermitian operator $R$ it holds that 
$||\Pi^{\perp}(R)||_{(\textrm{Tr})} = 2||P_{1}RP_{2}||_{(\textrm{Tr})}$. 
If we combine this with Eq.~(\ref{sdfnb}) we find that 
$2A_{(\textrm{Tr})}\boldsymbol{(}\Phi(\rho)\boldsymbol{)} 
= ||\Pi^{\perp}\circ\Phi(\rho)||_{(\textrm{Tr})} 
\leq ||\Pi^{\perp}(\rho)||_{(\textrm{Tr})} 
= 2A_{(\textrm{Tr})}(\rho)$, which proves the proposition.
$\Box$

Let $\boldsymbol{\mathcal{L}}= (\mathcal{L}_{1},\mathcal{L}_{2})$ 
be a decomposition of $\mathcal{H}$.
\begin{itemize}
\item For a channel $\Phi$ that is SP with respect to $\boldsymbol{\mathcal{L}}$ 
it holds that 
$A_{(\textrm{Tr})}^{\boldsymbol{\mathcal{L}}}\boldsymbol{(}\Phi(\rho)\boldsymbol{)}
\leq A_{(\textrm{Tr})}^{\boldsymbol{\mathcal{L}}}(\rho)$, 
for every density operator $\rho$ on $\mathcal{H}$.
\item For every unitarily invariant norm measure $A_{u}$ it holds that 
$A_{u}^{\boldsymbol{\mathcal{L}}}\boldsymbol{(}\Phi(\rho)\boldsymbol{)}
\leq A_{u}^{\boldsymbol{\mathcal{L}}}(\rho)$, for every density operator 
$\rho$ on $\mathcal{H}$, and every channel $\Phi$ that is LSP 
with respect to $\boldsymbol{\mathcal{L}}$.
\end{itemize}
Concerning the proofs of these statements note that we have already 
shown in Sec.~\ref{chin} that all SP channels satisfy the condition in 
Eq.~\ref{compoigen}, and thus SP channels cannot increase $A_{(\textrm{Tr})}$. 
It remains to prove that LSP channels do not increase any unitarily 
invariant norm  measure.  From Proposition 30 in Ref.~\cite{Ann} we know 
that if $\Phi$ is LSP with respect to $\boldsymbol{\mathcal{L}}$ then 
it follows that $P_{1}\Phi(\rho)P_{2} = V\rho W^{\dagger}$, 
where $V = \sum_{k}c_{1,k}V_{k}$ and $W = \sum_{l}c_{l}W_{l}$, 
and where 
$\sum_{k}V_{k}^{\dagger}V_{k} 
= P_{1}$,  $\sum_{k}V_{k}^{\dagger}V_{k}
 = P_{2}$,  $\sum_{k}|c_{1,k}|^{2}\leq 1$, 
and $\sum_{l}|c_{2,l}|^{2}\leq 1$. 
Moreover, $P_{1}V_{k}P_{1} = V_{k}$ and $P_{2}W_{k}P_{2} = W_{l}$. 
From these conditions it follows that $||V||_{(1)} \leq 1$ and $||W||_{(1)}\leq 1$. 
By using Proposition IV.2.4 in Ref.~\cite{Bhatia} we find that  
$||VP_{1}\rho P_{2}W^{\dagger}||_{(k)} \leq 
||V||_{(1)}||P_{1}\rho P_{2}||_{(k)}||W^{\dagger}||_{(1)}\leq 
||P_{1}\rho P_{2}||_{(k)}$. Thus,  we can conclude that 
$A_{(k)}\boldsymbol{(}\Phi(\rho)\boldsymbol{)} \leq A_{(k)}(\rho)$. 
Since this holds for all $k$, it follows, according to Proposition \ref{impl}, 
that  $A_{u}\boldsymbol{(}\Phi(\rho)\boldsymbol{)}\leq A_{u}(\rho)$ 
for all unitarily invariant norm measures.

In view of the results in this section there appears to be a  
correspondence between the class of induced measures and 
the class of invariant norm measures, and  especially between 
$A_{S}$ and $A_{(\textrm{Tr})}$, as is evident from Propositions 
\ref{nessuff} and \ref{nessuffigen}. 
However, it is far from clear to what extent this indicates a  profound 
relation between these measures. 
We finally note that  Propositions \ref{nessuff} and \ref{nessuffigen} 
shows that SP channels are limited in their capacity to create 
superposition since they cannot increase $A_{S}$ and $A_{\textrm{Tr}}$. 
However, as seen in Sec.~\ref{decay}, the other 
Ky-Fan norm measures may increase under the action of SP channels.

\section{\label{decay}Atom undergoing relaxation in an interferometer}
To illustrate the theory we consider models of an atom undergoing 
relaxation while propagating in a Mach-Zehnder interferometer. 
We apply the Ky-Fan norm measures to follow the evolution of the 
degree of superposition between the two paths.
We assume that the atom has $N$ relevant energy eigenstates 
spanning $\mathcal{H}_{I}$, e.g., electronic states, or nuclear spin 
states in an external magnetic field. 

To model the relaxation we let the total density operator $\rho$, 
describing both the internal and spatial degree of freedom, 
evolve according to a time-independent master equation that can be 
written on the Lindblad form \cite{Lindblmas},
\begin{eqnarray}
\label{Lindbladekv}
\frac{d}{dt}\rho & = & F(\rho),\nonumber \\
F(\rho) & = &  -i[H\otimes \hat{1}_{s},\rho]
 + \sum_{k}L_{k}\rho L_{k}^{\dagger}\nonumber\\
                 & & -\frac{1}{2}\sum_{k}L_{k}^{\dagger}L_{k}\rho 
- \frac{1}{2}\sum_{k}\rho L_{k}^{\dagger}L_{k},
\end{eqnarray}
where $H$ is the Hamiltonian of the atom, $\hat{1}_{s}$ is the identity 
operator on $\mathcal{H}_{s}$, and $L_{k}$ are the Lindblad 
operators. For convenience we assume $\hbar = 1$ and that energies 
are measured in units of some reference energy $E$, 
and likewise the dimensionless parameter $t$ measures time in units of $E^{-1}$.
The Lindblad form guarantees  that the evolution 
$\rho(t_{1}) = \Phi_{t_{1}-t_{0}}\boldsymbol{(}\rho(t_{0})\boldsymbol{)}$, 
for $t_{1}\geq t_{0}$, is such that the one-parameter family of dynamical 
maps $\Phi_{s}$ are channels \cite{Lindblmas}. 

We first note that there is a sufficient condition for the dynamical maps 
$\Phi_{s}$ of Eq.~(\ref{Lindbladekv}) to be SP with respect to the 
decomposition $(\mathcal{L}_{1},\mathcal{L}_{2})$. 
One can see that if $\Tr[P_{1}F(\rho)] = 0$ for all density operators $\rho$, 
then $\frac{d}{dt}\Tr(P_{1}\rho) =0$, which implies that all dynamical maps 
$\Phi_{s}$ are SP. A sufficient condition for $\Tr[P_{1}F(\rho)] = 0$ in 
Eq.~(\ref{Lindbladekv}) is that 
$\Pi(L_{k}) = L_{k}$.

We first consider two examples that admit analytical solutions.
Let $H=\hat{0}$ and assume the Lindblad operators  
$L_{k} = \sqrt{g}|e_{1}\rangle\langle e_{k}|\otimes\hat{1}_{s}$, 
for $k= 1,2,\ldots,N$, where $\{|e_{k}\rangle\}_{k=1}^{N}$ is an orthonormal 
eigenbasis of $\mathcal{H}_{I}$. Hence, all states relax to the ground state 
(a slight misnomer since $H=\hat{0}$) at the same rate.
The internal degree of freedom of the atom is prepared in the maximally mixed state 
(e.g., by letting it equilibrate at a sufficiently high temperature), and apply a 
50-50 beam-splitter to obtain the state 
$\rho(0) = \hat{1}_{N}\otimes|\psi\rangle\langle\psi|/N$, 
where $|\psi\rangle = (|1\rangle + |2\rangle)/\sqrt{2}$. 
Given this initial state we let the system evolve according to the master equation 
in Eq.~(\ref{Lindbladekv}), and thus propagate the atom at a sufficiently low 
temperature to allow relaxation to the ground state. (Again, this description is not 
entirely appropriate since $H=\hat{0}$, 
but is more adequate when we later let $H$ be nonzero.)  
By solving the master equation one finds that 
$A_{(1)}\boldsymbol{(}\rho(t)\boldsymbol{)} =  1/(2N) + (1-e^{-gt}) (N-1)/(2N)$, 
$A_{(k)}\boldsymbol{(}\rho(t)\boldsymbol{)} = 1/2-e^{-gt}[1/2-k/(2N)]$, 
for $k=2,\ldots,N$. Especially, we find that 
$A_{(\textrm{\Tr})}\boldsymbol{(}\rho(t)\boldsymbol{)}
 \equiv A_{(N)}\boldsymbol{(}\rho(t)\boldsymbol{)} = 1/2$.
 Hence, $A_{(\textrm{Tr})}$ is constant, 
which is consistent with the results in Sec.~\ref{chno}. 
It is to be noted that SP channels in general require nonlocal resources 
for their implementation, thus the increasing superposition as measured 
by the other Ky-Fan norm measures should not come as a surprise. 
As seen, all the Ky-Fan norm measures approach their maximal value $1/2$, 
which we know from Sec.~\ref{secbound} can be attained only if the total state is pure.
This is indeed the case since the internal state approaches the ground state, 
and does so in a manner that does not disrupt the superposition between the two paths.

For the second example we still assume $H = \hat{0}$, but let 
$L_{1:k} =  \sqrt{g}|e_{1}\rangle\langle e_{k}|\otimes|1\rangle\langle 1|$, 
and $L_{2:k} =  \sqrt{g}|e_{1}\rangle\langle e_{k}|\otimes|2\rangle\langle 2|$, 
for $k= 1,2,\ldots,N$, constitute the set of Lindblad operators.
In this case the superposition measures decay exponentially as 
$A_{(k)}\boldsymbol{(}\rho(t)\boldsymbol{)} = ke^{-gt}/(2N)$, $k=1,2,\ldots,N$.

\begin{figure}
\includegraphics[width = 7.5cm]{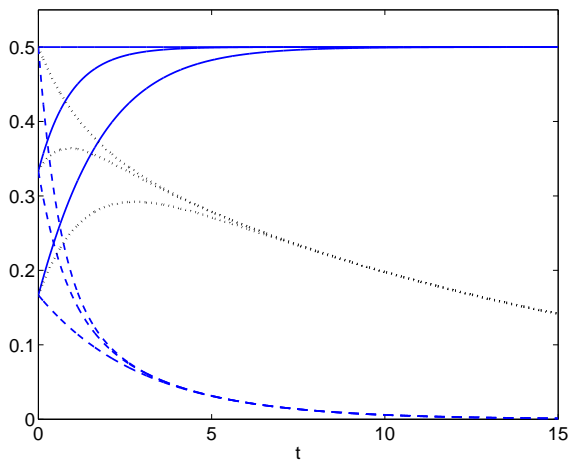}
\caption{\label{fig:relax} (Color online) Evolution of the Ky-Fan norm measures 
of the solution of a master equation describing an atom undergoing relaxation 
in the two paths of a Mach-Zehnder interferometer. 
The superposition measures are calculated with respect to the two paths. 
Counted from the top and down, the solid curves depict 
$A_{(3)}\boldsymbol{(}\rho(t)\boldsymbol{)}
\geq  A_{(2)}\boldsymbol{(}\rho(t)\boldsymbol{)}
\geq A_{(1)}\boldsymbol{(}\rho(t)\boldsymbol{)}$, 
where $\rho(t)$ is the solution of the master equation  
 in  Eq.~(\ref{Lindbladekv}) where the superoperator $F_{1}$ (the ``nonlocal" relaxation) is defined 
with respect to the Lindblad operators in Eq.~(\ref{vari1}). 
For convenience, the master equation is formulated such that $t$ is a 
dimensionless parameter.
The three dashed curves depict the evolution of the same superposition measures,  
but for a master equation with the superoperator $F_{2}$ (the ``local" relaxation)   
determined by the 
Lindblad operators in Eq.~(\ref{vari2}).
The three dotted lines gives the evolution resulting from the superoperator 
$F_{3} = 0.8 F_{1} + 0.2 F_{2}$, i.e., a convex combination of the two previous 
master equations. In all cases the internal  input state of the atom is maximally mixed. 
The atom is put in superposition between the two paths using a 50-50 beam-splitter. 
The Hamiltonian and the coefficients $g_{kk'}$ in Eqs.~(\ref{vari1}) and 
(\ref{vari2}) are chosen randomly, but are the same for all three cases. As seen, 
the local relaxation model appears to remove all superposition, while for the 
nonlocal relaxation all the Ky-Fan norm measures appear to approach 
their maximal value $1/2$. For the intermediate case $F_{3}$, 
two of the Ky-Fan norm measures initially increase, 
thus revealing the nonlocal nature of this case, since the corresponding channels
 cannot be LSP.}
\end{figure}

To a obtain a slightly more realistic model we assume a nonzero Hamiltonian 
and arbitrary decay rates between all eigenstates. 
We consider the superoperator $F_{1}$ as in Eq.~(\ref{Lindbladekv}) 
with the Lindblad operators 
\begin{equation}
\label{vari1}
L_{kk'}= \sqrt{g_{kk'}}|e_{k}\rangle\langle e_{k'}|\otimes\hat{1}_{s},
\quad 1\leq k\leq k'\leq N,
\end{equation}
 where $|e_{k}\rangle$ are the eigenstates of the Hamiltonian ordered 
in increasing energy. 
Figure \ref{fig:relax} shows the result of a numerical calculation, 
where the solid curves correspond to the evolution of the Ky-Fan norm 
measures $A_{(3)}$, $A_{(2)}$, and $A_{(1)}$, counted from the top and downward. 
(For the sake of clarity we consider a three-level system.)  
The eigenvalues of the Hamiltonian $H$ and the coefficients $g_{kk'}$ 
have been selected randomly.
We next consider the master equation with superoperator $F_{2}$, 
obtained from the Lindblad operators 
\begin{eqnarray}
\label{vari2}
L_{1:kk'} & = &\sqrt{g_{kk'}}|e_{k}\rangle\langle e_{k'}|\otimes|1\rangle\langle 1|,\nonumber\\
L_{2:kk'} & = &\sqrt{g_{kk'}}|e_{k}\rangle\langle e_{k'}|\otimes|2\rangle\langle 2|,
\end{eqnarray}
for $1\leq k\leq k'\leq N$.
 We use the same Hamiltonian, coefficients $g_{kk'}$, and initial state as 
in the previous example. In FIG.~\ref{fig:relax} the evolution of the 
superposition measures correspond to the dashed curves.
Finally, we consider the master equation with the superoperator 
$F_{3} = 0.8F_{1} + 0.2F_{2}$, i.e., a convex combination of the two previous
 master equations. We use the same initial state as before, and obtain the 
evolution of the three Ky-Fan norm measures as depicted in 
FIG.~(\ref{fig:relax}) with dotted lines.
 
As seen by the examples, there are two extreme cases: the ``nonlocal" relaxation 
where all the Ky-Fan norm measures approach the maximal value $1/2$, 
and the ``local" relaxation where they approach the minimal value $0$. 
One can show that the master equation of the local relaxation $F_{2}$ 
gives dynamical mappings that are LSP, 
while the nonlocal relaxation gives dynamical mappings that are general SP channels.
For master equations there is an implicitly assumed environment causing 
the nonunitary evolution. In the nonlocal case the master equation does 
not distinguish the path states, which we may interpret as the environment 
being insensitive to the path state of the atom. 
That there is no path information stored in the environment seems consistent 
with the nonvanishing superposition. In the case of local relaxation, however, 
local environments distinguish the paths and should reasonably cause decay 
of the superposition, as we indeed see is the case in FIG.~\ref{fig:relax}.
In the third case, with $F_{3}$, some of the dotted curves in FIG.~(\ref{fig:relax}) 
do increase beyond 
their initial value. It follows (by the results in Sec.~\ref{chno}) that the 
intermediate case (corresponding to $F_{3}$) cannot be LSP and thus 
represents a nonlocal relaxation. To summarize these examples, 
it appears as if the Ky-Fan norm measures to some extent reflect 
the locality or nonlocality of the relaxation process. 

One might wonder to what extent the nonlocal decay model is realistic. 
If the internal degree of freedom is a nuclear spin in an external magnetic field, 
and we assume that the system relaxates via spontaneous emission, 
the small energy splitting results in a long wavelength of the emitted photons.  
Moreover,  these states have a very long life time in general, 
which suggests a significant coherence length of the emitted radiation. 
If both these length scales are much larger than the separation of the two 
paths one might speculate  that the decay process does not ``notice" 
the path difference, and thus be nonlocal in the above sense. 
(There may, however, be other decoherence mechanisms that cause localization.)  
Along these lines one could consider experiments in order to test to what degree, 
and on what length scales, 
different relaxation mechanisms of various excited internal degrees of 
freedom do localize an atom.

In the above discussions we have interpreted the two paths of the interferometer 
as two spatial modes. However, we could also consider time binning 
(see, e.g., Ref.~\cite{Brendel} and a discussion in Ref.~\cite{OiJA}), 
two interfering decay channels, or some other degree of freedom. 
It is only required that there is no significant transition between the orthogonal 
subspaces corresponding to the two ``paths".

\section{\label{concl}Conclusions}
We introduce the concept of superposition measures with respect to given 
orthogonal decompositions of the Hilbert space of a quantum system, 
which can be regarded as an analogue to entanglement measures with 
respect to decompositions into subsystems. 
By a second quantization of the system, superposition can be regarded 
as entanglement, and  thus makes it possible to construct superposition 
measures using entanglement measures.
 We find the superposition measures induced by relative entropy of 
entanglement and of entanglement of formation, and obtain a 
decomposition formula of the relative entropy of entanglement for certain 
classes of states. 
We furthermore consider a class of superposition measures based on 
unitarily invariant operator norms. 
Especially we consider measures derived from Ky-Fan norms, 
and show that these can be obtained operationally using interferometric techniques. 
We show that the Ky-Fan norm based measures are bounded by predictability, 
similarly as for the standard visibility in interferometry \cite{Englert, Durr}.
 We furthermore consider the superposition measures under the action of 
subspace preserving and local subspace preserving channels \cite{Ann}, 
and show that these channels cannot increase the superposition with respect 
to certain superposition measures. We illustrate the theory with models of an 
atom undergoing relaxation while propagating in a Mach-Zehnder interferometer.
We consider ``local" and ``nonlocal" relaxation models,   
and monitor this difference by the evolution of the superposition measures.

One could consider to extend the ideas presented here by defining
 a ``superposition cost" by inducing 
it from the entanglement cost \cite{cost1}. Similarly, one can  induce a 
``distillable superposition" from distillable entanglement \cite{mseerr}. 
It would be interesting to find formulations of these induced superposition measures 
directly in terms of the subspace decompositions of $\mathcal{H}$, 
similarly as done here for the relative entropy of superposition and superposition 
of formation, 
rather than via the indirect definition using the second quantization.  

The quantitative approach to superposition introduced here may facilitate 
the development of interferometry as a probing technique of processes in 
physical systems \cite{Ann, Oi, JA, OiJA, Xiang}. Especially, 
one could consider to use the superposition measures to 
systematically construct interferometric channel 
measures in the spirit of Refs.~\cite{Oi, JA, OiJA}.

One could furthermore consider to combine the superposition measure 
concept introduced here, and the channel measures in  
Refs.~\cite{Oi, JA, OiJA}, with the perspective put forward in Ref.~\cite{Braun}.  
The interference measure introduced in Ref.~\cite{Braun} quantifies 
how sensitive the outcomes of measurements in the computational basis 
are to phase changes of input superpositions to a quantum  process. 
It does not seem unreasonable that there exist relations between 
this interference measure and the change of superposition caused 
by the channel. Such a comparison would  benefit from a generalization
 of the measure put forward in Ref.~\cite{Braun} to the general type of 
subspace decompositions considered here. 

Superposition measures may also be useful to the efforts to find meaningful 
generalizations of geometric phases and holonomies \cite{berry84,wilczek84} 
to mixed states and open systems 
\cite{Ell, ME2, Sjo, Pei, Tong, Cha, Sar, ME, rezakhani06}. 
One may for example note that in Ref.~\cite{Uinf} the Uhlmann holonomy 
\cite{Uhl} have been formulated within a framework closely related to the 
type of subspace structures considered here. 

\acknowledgements
The author wishes to thank the Swedish Research Council for
financial support and the Centre for Quantum Computation at DAMTP,
Cambridge, for hospitality.


\begin{thebibliography}{99}
\bibitem{comment} There exists a relation between separable channels and 
relative entropy of entanglement that reminds of Proposition \ref{nessuff}. 
A channel $\Phi$ on $\mathcal{H}_{A}\otimes\mathcal{H}_{B}$  is called 
separable if it has a Kraus representation $\{A_{k}\otimes B_{k}\}_{k}$ \cite{Rains,BenDiVi}.  
By combining some results in Refs.~\cite{entangmes} and \cite{CiDu} one can 
 shown that $\Phi$ is separable if and only if 
$E_{S}\boldsymbol{(}[\Phi\otimes I_{A'}\otimes I_{B'}](\rho)\boldsymbol{)}\leq E_{S}(\rho)$, 
for all $\rho$ on 
$\mathcal{H}_{A}\otimes \mathcal{H}_{A'}\otimes\mathcal{H}_{B}\otimes\mathcal{H}_{B'}$, 
where $\mathcal{H}_{A'}$ and $\mathcal{H}_{B'}$ are copies of $\mathcal{H}_{A}$ 
and $\mathcal{H}_{B}$, respectively. The extension to a larger Hilbert space 
is necessary for this relation to be valid 
(e.g., $E_{S}\boldsymbol{(}\Phi(\rho)\boldsymbol{)} = E_{S}(\rho)$ if $\Phi$ 
is the swap channel \cite{CiDu}). Hence, the analogy with Proposition  \ref{nessuff} 
is not complete, since the latter does not require such extensions.
\end{thebibliography}
\end{document}